\documentclass{aa}
\usepackage{bibtex/natbib}
\usepackage{xspace}
\usepackage{amssymb}
\usepackage{amsmath}
\usepackage{graphicx}
\usepackage{verbatim}

\def\Sa{Hit \& Stick (S1)}
\def\Sb{Sticking through surface effects (S2)}
\def\Sc{Penetration (S3)}
\def\Sd{Mass transfer (S4)}
\def\Ba{Bouncing with compaction (B1)}
\def\Bb{Bouncing with mass transfer (B2)}
\def\Fa{Fragmentation (F1)}
\def\Fb{Erosion (F2)}
\def\Fc{Fragmentation with mass transfer (F3)}

\def\pp{\textit{`pp'}}
\def\pP{\textit{`pP'}}
\def\cc{\textit{`cc'}}
\def\cC{\textit{`cC'}}
\def\pc{\textit{`pc'}}
\def\pC{\textit{`pC'}}
\def\cp{\textit{`cp'}}

\begin{document}

\title{The outcome of protoplanetary dust growth:\\
pebbles, boulders, or planetesimals?  \thanks{This paper is dedicated to the memory of our dear friend and colleague Frithjof Brauer (14th March 1980 - 19th September 2009) who developed powerful models of dust coagulation and fragmentation, and thereby studied the formation of planetesimals beyond the meter size barrier in his PhD thesis. Rest in peace, Frithjof.}}
\subtitle{II. Introducing the bouncing barrier}
\authorrunning{A. Zsom et al.}
\titlerunning{The outcome of protoplanetary dust growth:\\
pebbles, boulders, or planetesimals? II.}
\author{A. Zsom $^{1}$,  C. W. Ormel  $^{1}$, C. G\"{u}ttler $^{2}$, J. Blum $^{2}$,  C. P. Dullemond  $^{1}$}
\institute{Max-Planck-Institute f\"ur Astronomie, K\"onigstuhl 17,
  D-69117 Heidelberg, Germany. Email: \texttt{zsom@mpia.de}
  \and
Institut f\"{u}r Geophysik und extraterrestrische Physik, Technische 
Universit\"{a}t Braunschweig, Mendelssohnstr. 3, 38106 Braunschweig, 
Germany. Email: \texttt{c.guettler@tu-bs.de}}
\date{\today}

  \abstract
  % context heading (optional)
  % {} leave it empty if necessary  
   {The sticking of micron sized dust particles due to surface forces in circumstellar disks is the first stage in the production of asteroids and planets. The key ingredients that drive this process are the relative velocity between the dust particles in this environment and the complex physics of dust aggregate collisions.}
  % aims heading (mandatory)
   {Here we present the results of a collision model, which is based on laboratory experiments of these aggregates. We investigate the maximum aggregate size and mass that can be reached by coagulation in protoplanetary disks.}
  %{later}%
  % methods heading (mandatory)
   {We use the results of laboratory experiments to establish the collision model (G\"{u}ttler et al. (2009)). The collision model is based on some necessary assumptions: we model the aggregates as spheres having compact and porous `phases' and a continuous transition between these two. We apply this collision model to the Monte Carlo method of Zsom \& Dullemond (2008) and include Brownian motion, radial drift and turbulence as the sources of relative velocity between dust particles.}
  %{later}%
  % results heading (mandatory)
  %{later}%
  {We model the growth of dust aggregates at 1 AU at the midplane at three different gas densities. We find that the evolution of the dust does not follow the previously assumed growth-fragmentation cycles. Catastrophic fragmentation hardly occurs in the three disk models. Furthermore we see long lived, quasi-steady states in the distribution function of the aggregates due to bouncing. We explore how the mass and the porosity change upon varying the turbulence parameter and by varying the critical mass ratio of dust particles. Upon varying the turbulence parameter, the system behaves in a non-linear way and the critical mass ratio has a strong effect on the particle sizes and masses. Particles reach Stokes numbers of roughly $10^{-4}$ during the simulations.}
  % conclusions heading (optional), leave it empty if necessary 
   {The particle growth is stopped by bouncing rather than fragmentation in these models. The final Stokes number of the aggregates is rather insensitive to the variations of the gas density and the strength of turbulence. The maximum mass of the particles is limited to $\approx$ 1 g (chondrule sized particles). Planetesimal formation can proceed via the turbulent concentration of these aerodynamically size-sorted chondrule-sized particles. }
   
     %Particles initially grow by sticking mechanisms which is sometimes followed by a transition regime where a variety of sticking and bouncing collisions occur. The evolution of the particles is then halted by bouncing which is the dominating collision type during this phase. At this stage the mass of the particles decreases in time due to a low probability of breakage while bouncing. 
   

\maketitle

\section{Introduction}
In the core accretion paradigm of planet formation (\cite{Mizuno1980}; \cite{Pollack1996}) planets are the outcome of an accretion process that starts with micron-size dust grains and covers 40 magnitudes in mass. It can be divided into three stages. The first stage of the formation of rocky planets and the rocky cores of gas giant planets starts with the coagulation of dust in the protoplanetary disks surrounding many pre-main-sequence stars (\cite{Safronov1969}, \cite{Weidenschilling1993}, \cite{Blum2008}). The next stage of planet formation is the formation of protoplanetary cores from the planetesimals. The idea is that the kilometer size planetesimals are so large, that gravity starts to take over and leads to the gravitational agglomeration of these bodies to rocky planets. This scenario was studied already by \cite{Safronov1969}, and has since been modeled using numerical methods by \cite{Weidenschilling1980}, \cite{Nakagawa1983}, \cite{Mizuno1988}, \cite{Schmitt1997}, \cite{Wetherill:1990p85}, \cite{Nomura2006}, \cite{Garaud2004}, \cite{Tanaka2005} and several more authors. These models solve for the size distribution of dust aggregates in the disk as a function of time, and investigate if, where and how larger dusty bodies form, and how long that takes. Finally, in the third stage, gas accretes onto these protoplanets forming giant planets or -- in the absence of gas -- gravitational encounters between these protoplanets result in a chaotic, giant impact phase, until orbital stability has been achieved (\cite{Chambers2001}; \cite{Kokubo2006}; \cite{Thommes2008}).\\

In this study, we focus on the first phase and address the fundamental question of how effective dust growth by surface forces really is; that is, how big do particles become by simple sticking processes only.  It is known that initially, for micron size grains, the growth is driven by Brownian motion. This typically leads to slow collisions and forms aggregates of fractal structure (\cite{Kempf1999}; \cite{Blum1996}). In the current picture of dust growth, as these aggregates grow, at some point the growth will leave the fractal regime, and collisions will start to lead to compaction and breaking of the aggregates (\cite{Blum2000}), embedding of small bodies into larger aggregates (leading to `filling up' of these larger aggregates and compaction due to the force of the collision (\cite{Ormel:2007p93}). As the size of the dust aggregates increases, differential vertical settling (\cite{Safronov1969}), radial drift (\cite{Whipple1972}) and turbulence (\cite{Voelk1980}; \cite{Mizuno1988}; \cite{Ormel:2007p92}) will become important new mechanisms driving relative velocities between aggregates. The increasing relative velocities caused by these mechanisms will at least partly compensate the lower collision probability due to lower surface-over-mass ratio of large aggregates. When the aggregates grow to sizes of millimeter to meter, however, the sticking efficiency drops strongly (e.g. \cite{Blum1993}) and the relative velocities become so large that aggregates can fragment (\cite{Blum2008}, so called `fragmentation barrier'). Another hurdle that the particles have to circumvent is the `drift barrier' (\cite{Weidenschilling1977}), namely that millimeter, centimeter sized particles are lost to the star due to radial drift in a short timescale. Recently, \cite{Okuzumi2009} pointed out the existence of a `charge barrier', which possibly halts the particle
 growth already at an early stage of fractal aggregates. Despite many years of efforts, it is not known if the coagulation process can overcome these barriers. These barriers have been and still are the main open question of the initial stages of planet formation: the growth from dust to planetesimals. \\

Several mechanisms have been proposed to overcome this problem, among which are the trapping of dust in vortices (\cite{Barge1995}; \cite{Klahr1997}, \cite{Lyra2009a}), trapping of decimeter-sized boulders in turbulent eddies and the subsequent gravitational collapse of swarms of these trapped boulders (\cite{Johansen:2007p65}), the trapping of particles in a pressure bump caused by the evaporation front of water (\cite{Kretke2007}; \cite{Brauer2008b}) and many more scenarios. However, the correct modeling of any of these scenarios requires the detailed knowledge of the collisional physics, and these models have so far relied either on simplified input phyisics or on simplified initial conditions.

Because of their complexity, collisional evolution models have to make simplifying assumption concerning the outcome of dust aggregate collisions, for example that collisions always result in sticking, or otherwise use simple recipes for the collisional outcome.  Ideally, one requires to know the detailed outcome of every collision.  But modelling this microphysics within an evolution model is simply unpractical. There are computer programs that model such individual collisions in detail (e.g. \cite{Dominik:1997p89}, \cite{Suyama2008}; Geretshauser et al, in prep.), but each model collision takes anywhere from hours to weeks to run on a computer. They are therefore not practical to use at run-time in a model that computes the overall time-dependent evolution of the dust size distribution inside protoplanetary disks. Moreover, such collision models themselves often depend on poorly known input physics. 

Another approach to obtain the collisional outcome of dust aggregates is to model these collision in the laboratory. From the many experiments that have as of now been performed a picture emerges of the outcome of dust aggregate collision under a variety of conditions in the {protoplanetary disk (PPD). In a companion paper (G\"{u}ttler et al. (2009), henceforth Paper I), we have collected data from over 19 experiments, and constructed a set of formulae that reasonably well describe the outcomes of these collisions in such a way that they can be used as input for models that address the temporal evolution of the dust size distribution.

In this paper we will directly rely on the outcome of these laboratory experiments for modeling the dust aggregate size distribution. As described in Paper I, we have produced a mapping of all available collision experiments regarding silicate-like particles. The velocity range of these experiments is also sufficiently wide to cover various disk models which roughly correspond to the conditions at 1 AU in the PPD. For details regarding the collisional mapping, we will refer to paper I, but we will summarize the elements of our new collision model in Sect. 3.1.

We build this collision kernel into a Monte Carlo code for modeling the size- and porosity distribution of dust in a protoplanetary disk (\cite{Zsom2008}, hereafter ZsD08). The outcome of our laboratory-driven dust coagulation model is hard to a priori predict since the key variables involved depend on a non-trivial interplay between the collision kernel (Paper I) and the velocity field.  We can, however, anticipate two scenarios. In the first, particle growth will proceed beyond the meter-size barrier, all the way to planetesimals. In the second scenario, growth will terminate at an intermediate size. In this case further growth to planetesimal sizes may proceed through concentration and subsequent gravitational collapse of these particles (\cite{Johansen:2007p65}, \cite{Cuzzi2008}). Thus, our model will provide the starting conditions for these concentration models. We do emphasize, however, that in this work we do not in any way 'optimize' the outcome by laboriously scanning all the parameter space or treating environments that may be more conducive for growth, like nebula pressure bumps or trapping of dust in vortices (\cite{Kretke2007}, \cite{Lyra2009a}). These are obvious expansions of our work. But by considering the sensitivity of a few key parameters (e.g., gas density, and turbulence strength) on the outcome of the growth process, we do obtain a picture of where the arrow of coagulation typically points to in protoplanetary environments: pebbles, boulders or planetesimals.\\

In this paper we describe the three nebulae models used in this work and the sources of relative velocity between the aggregates (Sect. \ref{sec:2}), how we build the coagulation/fragmentation model of Paper I into the Monte Carlo code (Sect. \ref{sec:impl}), and what these first results look like (Sect. \ref{sec:res}). We also test the sensitivity of the results with respect to variations in gas density, the velocity field, and other key model parameters. Section  \ref{sect:disc} reflect the importance of our result in the context of planetesimal formation and provide suggestions for future experiments. Finally, Sect. \ref{sec:sum} lists our main conclusions. 

\section{The nebulae model}
\label{sec:2}
\subsection{Disk models}
\label{sec:disks}
In this Section we briefly describe the disk models considered in this paper.

\paragraph{The low density model:}
Resolved millimeter emission maps of protoplanetary disks seem to indicate a shallow surface density profile (\cite{Andrews2007}): $\Sigma_g(r) \propto r^{-0.5}$. Systematic effects of some of their assumptions, such as the disk inclinations or the simplified treatment of the temperature distribution, may suggest somewhat steeper profiles. Therefore, \cite{Brauer2008a} adopted the following profile:
\begin{equation}
\Sigma_g(r)= 45 \frac{\mbox{ g}}{\mathrm{cm}^2} \left( \frac{r}{\mathrm{AU}} \right)^{-0.8}.
\end{equation}
Here we assumed that the central star is of solar mass, the disk extends from 0.03 AU until 150 AU and that the total mass of the disk is 0.01 M$_\odot$. Assuming that the pressure scale-height is $H_p=0.05\times r$ and the vertical structure is gaussian:
\begin{equation}
\rho_g (z,r) = \frac{\Sigma_g(r)}{\sqrt{2 \pi} H_p}\exp(-z^2/2H_p^2),
\end{equation}
the density at 1 AU in the midplane ($z=0$) is $2.4 \times 10^{-11}$ g cm$^{-3}$, approximately two orders of magnitude lower than the Minimum Mass Solar Nebulae (MMSN) value.

\paragraph{MMSN model: }
The Minimum Mass Solar Nebulae model (MMSN) was introduced by \cite{Weidenschilling1977a} and \cite{Hayashi1985}. From the present state of the Solar System today, it is possible to obtain a lower limit to the mass if the solar nebulae from which the planets were formed. The model assumes that the planets were formed where they are currently located (no migration included). It also assumes that all the solid material presented in the solar nebula had been incorporated in the planets. The loss of solid material  due to radial drift is not taken into account. Despite these uncertainties, the MMSN model is frequently used as a benchmark. The surface density of the MMSN disk is given by:
\begin{equation}
\Sigma_g(r)= 1700 \frac{\mbox{ g}}{\mathrm{cm}^2} \left( \frac{r}{\mathrm{AU}} \right)^{-1.5},
\end{equation}
which corresponds to a total disk mass of 0.01 M$_\odot$ contained between 0.4 and 30 AU (between the orbits of Mercury and Neptune). Assuming that the vertical structure of the gas follows a gaussian distribution, leads to a midplane density at 1 AU of $1.4\times 10^{-9}$ g cm$^{-3}$. 

\paragraph{The high density model:}
\cite{Desch2007} introduced a `revised MMSN model' by adopting the starting positions of the planets in the 'Nice' model of planetary dynamics (\cite{Tsiganis2005}) thus taking into account planetary migration. The model predicts that the solar system started out in a much more compact configuration and its surface density profile is given by:
\begin{equation}
\Sigma_g(r)= 5.1\times 10^4 \frac{\mbox{ g}}{\mathrm{cm}^2} \left( \frac{r}{\mathrm{AU}} \right)^{-2.2}.
\end{equation}
This model is consistent with a decretion disk which is being photoevaporated by the central star. Although the model of \cite{Desch2007} was defined for the outer solar system, we extrapolate the profile to 1 AU in order to cover a broad range of surface density values in our calculations. Assuming, as in the MMSN model, a gaussian vertical distribution, the density at 1 AU in the midplane is $2.7 \times 10^{-8}$ g cm$^{-3}$.\\

For simplicity, we adopt a midplane temperature of 200 K (isothermal sound speed of $c_s = 8.5\times 10^{4}$ cm s$^{-1}$) in all the three models.

\subsection{Relative velocities}
\label{sec:vrel}
We consider three sources for relative velocities between dust aggregates. These are Brownian motion, radial drift and turbulence. In the following, we discuss these sources.

\begin{figure*}
  \includegraphics[width=0.5\textwidth]{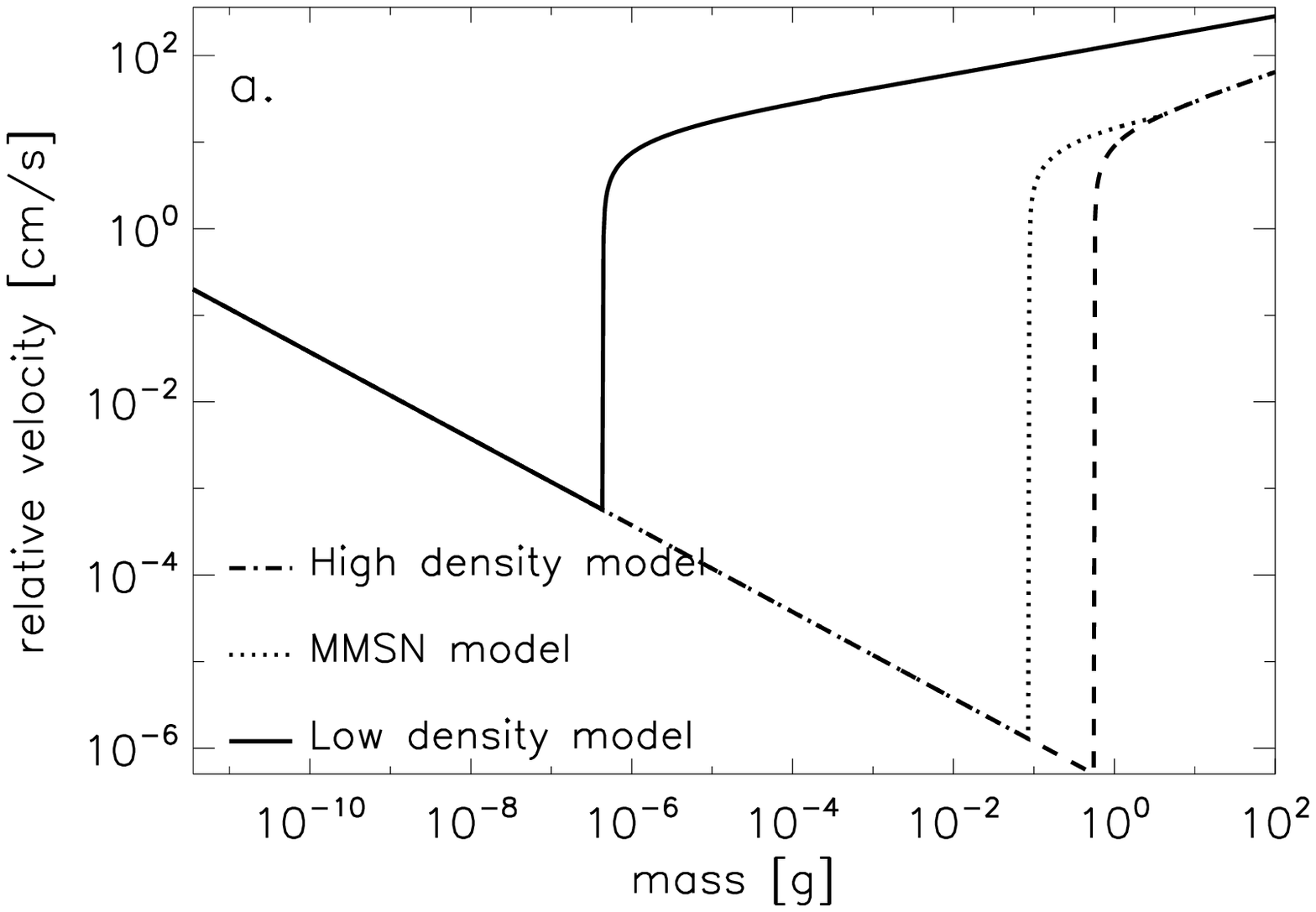}
  \includegraphics[width=0.5\textwidth]{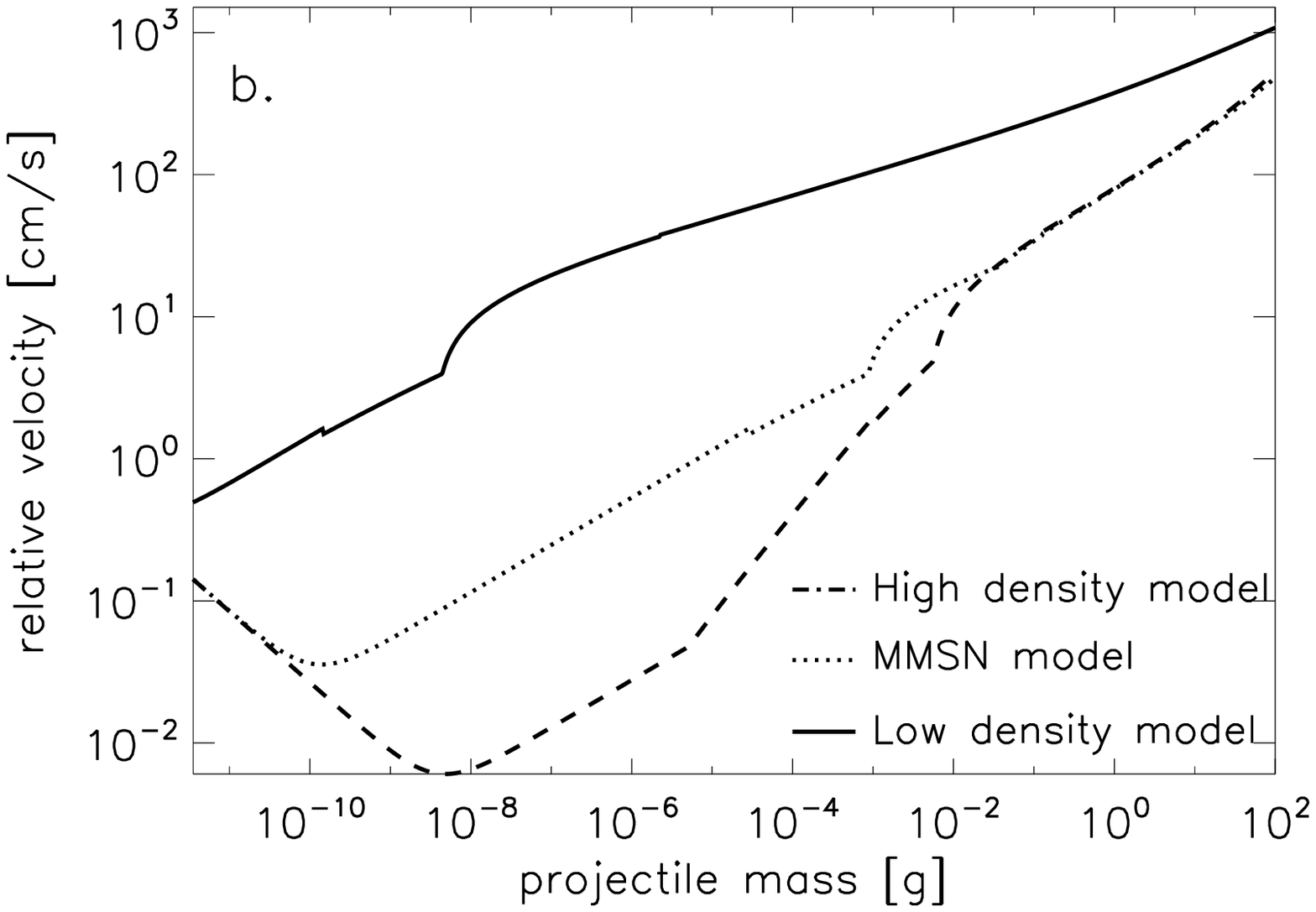}
  \caption{The combined relative velocities caused by Brownian motion, radial drift and turbulence for fluffy particles ($\Psi = 20$) in the three disk models for equal sized particles (a) and for different sized particles with a mass ratio of 100 (b). The solid line indicates the low density model of \cite{Brauer2008a}. Physical parameters of the disk: the distance from the central star is 1 AU, temperature is 200 K, the density of the gas is $2.4\times 10^{-11}$ g cm$^{-3}$, and the turbulence parameter, $\alpha=10^{-4}$. The dotted line represents the MMSN model. The density is $1.4 \times 10^{-9}$ g cm$^{-3}$, the other parameters are the same. The dashed line corresponds to the high density disk. The gas density is $2.7\times 10^{-8}$ g cm$^{-3}$.}
  \label{relcont}
\end{figure*}

The average relative velocity of two particles with mass $m_1$ and $m_2$ in a region of a disk with temperature $T$ due to Brownian motion is
\begin{equation}
 \Delta v_B (m_1,m_2) = \sqrt \frac{8kT(m_1+m_2)}{\pi m_1 m_2}.
\end{equation}
For micron sized particles, the relative velocity is of the order of 0.1 cm s$^{-1}$, but for cm sized particles this value drops several orders of magnitude. Therefore, Brownian motion is only effective for collisions between small particles during the initial stages of growth. Coagulation due to Brownian motion results in fluffy aggregates of fractal dimension around 2 and 3 (\cite{Blum1996}; \cite{Kempf1999}; \cite{Blum2000a}; \cite{Krause2004}). In practice there is no growth due to Brownian motion for aggregates larger than 100 micron. 

The second source for relative velocity is turbulence. Relative velocity of aggregates due to the random motion of turbulent eddies were calculated numerically by \cite{Voelk1980}, \cite{Mizuno1988} and \cite{Markiewicz1991}. We use the closed form expressions presented by \cite{Ormel:2007p92}. We assume that turbulence is parameterized by the \cite{Shakura1973} $\alpha$ parameter
\begin{equation}
\nu_T=\alpha c_s H_g,
\label{eq:nuT}
\end{equation}
where $\nu_T$ is the turbulent viscosity, $c_s$ is the isothermal sound speed and $H_g$ is the pressure scale height of the disk. The value of the $\alpha$ parameter reflects the strength of the turbulence in the disk. Typical values of $\alpha$ in this paper range between 10$^{-3}$ and 10$^{-5}$. The turbulent relative velocity is a function of the stopping times of the two colliding particles. The stopping time (or friction time) is the time the particle needs to react to the changes in the motion of the surrounding gas. As long as the radius of the particle is smaller than the mean free path of the gas ($a < \frac{9}{4}\lambda_{\mathrm{mfp}}$), the particle is in the Epstein regime, where the stopping time is (\cite{Epstein1924}):
\begin{equation}
t_{s} = t_{\mathrm{Ep}} = \frac{3 m}{4 v_{\mathrm{th}} \rho_g A},
\label{eq:ts1}
\end{equation}
where $m$ and $A$ are the mass and the cross section of the particle, $\rho_g$ and $v_{\mathrm{th}}$ are the gas density and the thermal velocity. At high gas densities, where the mean free path is low or in case of larger particles, the first Stokes regime applies and the stopping time is
\begin{equation}
t_s = t_{\mathrm{St}} = \frac{3 m}{4 v_{\mathrm{th}} \rho_g A} \times \frac{4}{9} \frac{a}{\lambda_{\mathrm{mfp}}}.
\label{eq:ts2}
\end{equation} 
In the first Stokes regime the stopping time is independent of the particle-gas relative velocity as well as the gas density. This regime can be used as long as the particle Reynolds number is smaller than unity. The particle Reynolds number calculated as (\cite{Weidenschilling1977}):
\begin{equation}
Re_p = \frac{2 a \Delta v_{\mathrm{pg}}}{\eta},
\end{equation}
where $\Delta v_{\mathrm{pg}}$ is the relative velocity between the particle and the gas, and $\eta$ is the gas viscosity. For particles outside the Epstein regime, it can be assumed that the systematic velocity (radial drift) dominates over the random velocities (turbulence); therefore, $\Delta v_{\mathrm{pg}} \approx v_D$, where $v_D$ is the drift velocity of the particle, defined in the next paragraph. The particle Reynolds number never exceeds unity in our simulations. Therefore, we do not include further Stokes regimes.

\begin{figure}
  \includegraphics[width=0.5\textwidth]{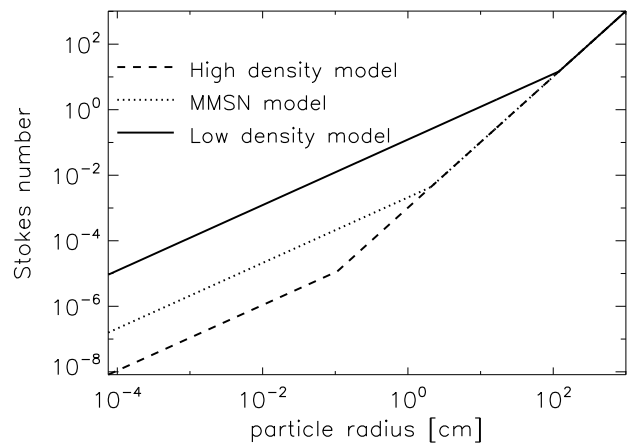}
   \caption{The Stokes number as a function of the particle radius in the three models. The parameters of the dust for all of the models are the following: monomer radius is $a_0 =0.75$ $\mu$m, material density is $\rho_0 = 2$ g cm$^{-3}$, and $\Psi = 1$.}
  \label{fig:ep-st}
\end{figure}

Radial drift also leads to relative velocities between aggregates. Radial drift ($v_D$) has two sources: drift of individual particles ($v_{d}$) and drift due to accretion processes of the gas ($v_{da}$), thus the total radial drift velocity is $v_D=v_{d}+v_{da}$.\\
The radial drift of individual dust aggregates with mass $m$ is (\cite{Weidenschilling1977})
\begin{equation}
v_{d} = -\frac{2 v_N}{St+1/St} ,
\end{equation}
where $St$ is the Stokes number of the aggregate ($St=t_s\Omega$, where $\Omega$ is the orbital frequency) and $v_N$ is the maximum radial drift velocity (\cite{Whipple1972}). 

The second part of the radial velocity is due to the accretion of the gas. This part of the radial velocity is calculated as follows (\cite{Kornet2001}):
\begin{equation}
v_{da} = \frac{v_{\mathrm{gas}}}{1+St^2},
\label{eq:drift}
\end{equation}
where $v_{\mathrm{gas}}$ is the accretion velocity of the gas (\cite{Takeuchi2002}). 

The relative velocity due to radial drift is then simply the difference between the radial velocity of particle 1 and particle 2. However, as the Stokes number of the aggregates is always smaller than $10^{-3}$ (see Sec. \ref{sec:res}), the second term of the radial velocity ($v_{da}$) can be safely neglected:
\begin{equation}
\Delta v_D = | v_{D1} - v_{D2}| \approx | v_{d1} -v_{d2}|.
\end{equation}
\\

This study uses two quantities to describe the porosity of the aggregates. The volume filling factor is:
\begin{equation}
\phi = V^*/V_{\mathrm{tot}}=(A^*/A)^{3/2},
\end{equation}
where $V^*$ is the volume occupied by the monomers and $V_{\mathrm{tot}}$ is the total volume of the aggregate, including pores, and $A$ and $A^*$ are the surface area equivalents of these quantities. In this way, the filling factors also enters the definition of the friction time (Eqs. \ref{eq:ts1} and \ref{eq:ts2}). The density of aggregates then follows as $\rho=\rho_0 \phi$, where $\rho_0 = 2$ g cm$^{-3}$ is the material density of the silicate. In this study we will also use the reciprocal parameter of the filling factor, which is denoted the enlargement parameter, $\Psi = \phi^{-1}$.\\

We illustrate the relative velocity between equal sized and different sized aggregates with $\Psi = 20$ ($\phi = 0.05$) in Fig. \ref{relcont} for the disk models considered in this work. Adopting a threshold (fragmentation) velocity of 1 m s$^{-1}$, the maximum particle size, which can be reached in the models are: 0.025 cm in the low density model, 1.4 cm in the MMSN model and 1.7 cm in the Desch model. The Stokes numbers of these particles are the same in all the three models, $4.7 \times 10^{-3}$. The constant fragmentation velocity of 1 m s$^{-1}$ is the typical velocity at which silicate particles will fragment (Birnstiel et al. (2009)). In our collision model this is not the case for all combinations of mass ratio and porosity (Paper I), but the m s$^{-1}$ threshold is still a useful proxy for the point where fragmentation processes will become important.

Figure \ref{fig:ep-st} shows the Stokes number as a function of particle radii in the three models. Initially, particles are in the Epstein regime, where the stopping time, thus the Stokes number, depends on the gas density. When the particles enter the Stokes regime, the stopping time becomes independent of the gas density (see Eq. \ref{eq:ts2}). One can see that particles in the Desch model are in the Stokes regime at Stokes number of $4.7 \times 10^{-3}$ (when the particles have relative velocities of 1 m s$^{-1}$), while the aggregates in the MMSN model are close to it, which explains why the maximum particle size is almost the same in these two models.

As discussed in \cite{Ormel:2007p92}, particles are initially in the `tightly coupled particle' regime, where the eddies are all of class I type, meaning that the turnover time of all eddies is longer than the friction time of the particles (\cite{Voelk1980}). A particle, upon entering a class I eddy, will therefore forget its initial motion and align itself to the gas motions of the eddy before the eddy decays or the particle leaves it. This regime is apparent in Fig. \ref{relcont}a and b. Different sized particles are in this relative velocity regime as long as their masses are less than $10^{-8}$ g in the low density model, $10^{-3}$ g in the MMSN model and $10^{-2}$ g in the high density model assuming fluffy particles ($\Psi = 20$). If the particles leave this regime and enter the `intermediate particle' regime, their relative velocity increases. This transition affects the particle evolution, as discussed in e.g. Sect \ref{sec:mmsn}.

\section{Collision model and implementation}
\label{sec:impl}
In this work we use a statistical or `particle in a box' method to compute the collisional evolution. That is, we assume that all particles are homogeneously distributed within a certain volume (the simulation volume). In reality however, the particles could leave the simulated volume or new particles could enter from outside due to radial drift or random motions (turbulence and Brownian motion). Since we do not resolve the spatial dependence of the aggregates, we will simply assume that local conditions hold during the run. The gas and dust densities are kept constant and particles cannot leave or enter the simulation volume (hereafter `local approach').

\subsection{Short overview of the collision model}

Many laboratory experiments on dust aggregate collisions have been performed in the past years, see \cite{Blum2008}. The growth begins as fractal growth and we use the recipe of \cite{Ormel:2007p93} to describe this initial stage. However, once aggregates have restructured into non-fractal, macroscopic aggregates (e.g. $\gtrsim100\;\mu$m), laboratory experiments show that the collisional outcomes become very diverse. In this regime, many new experiments were performed with dust aggregates consisting of 1.5 $\mu$m diameter SiO$_2$ monomers either with high porosity $\phi=0.15$ (\cite{Blum2004}), or intermediate porosity ($\phi=0.35$). Paper I compiled 19 experiments with different aggregate masses, collision velocities, and aggregate porosities.\\

\begin{figure*}
\centering
  \includegraphics[width=0.7\textwidth]{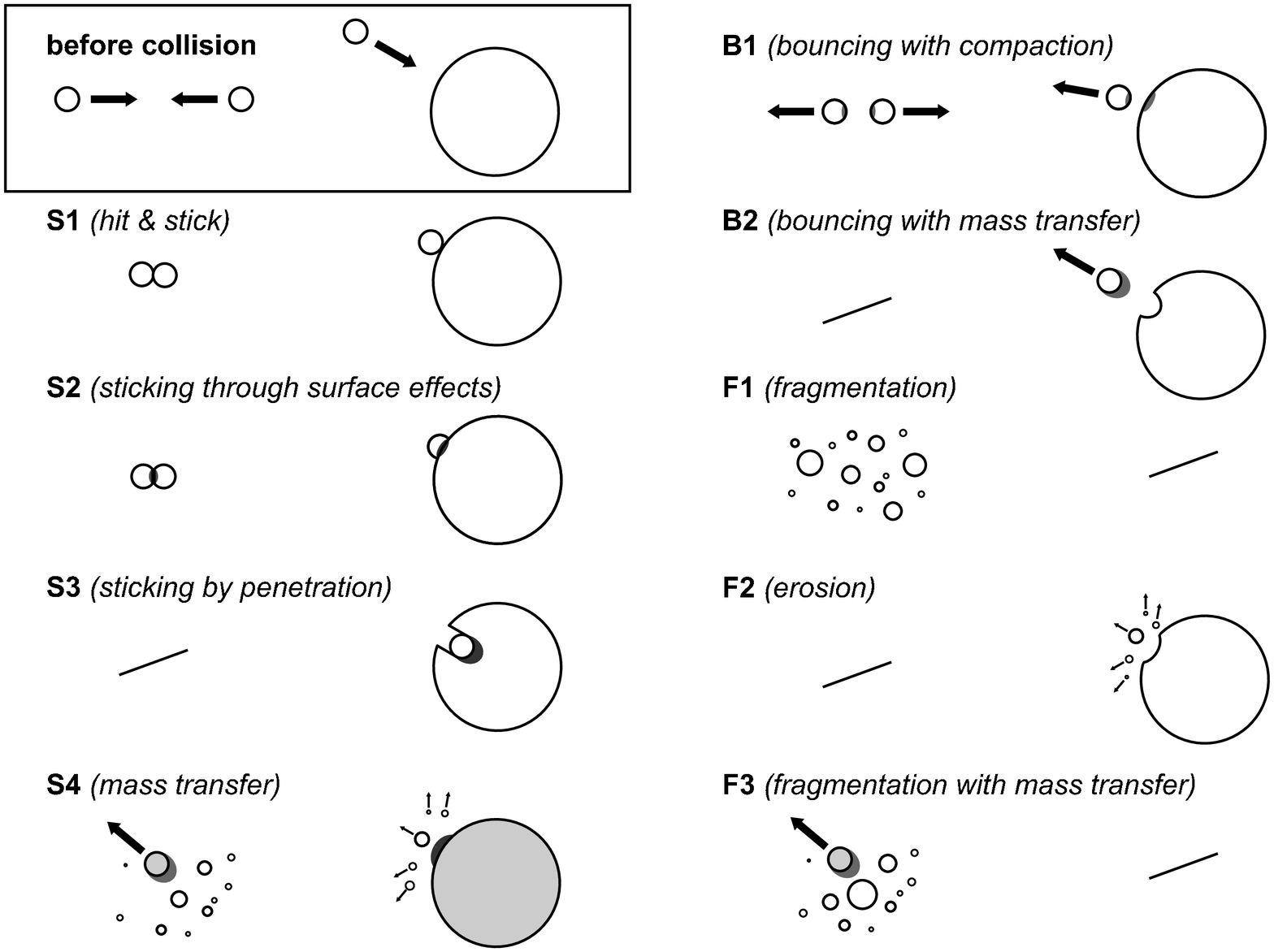}
  \caption{The collision types considered in this paper. We distinguish between similar sized and different sized particles. Some of the collision types only happens for one of the mass ratios. Grey color indicates that during the given collision type the particle is compact, or part of the mass will be compacted.}
  \label{coltypes}
\end{figure*}

From these experiments we have identified nine different collisional outcomes involving sticking, bouncing, or fragmentation (see Fig. \ref{coltypes}). The occurrence of these regimes mainly depends on aggregate masses and collision velocities. However, it also depends on the porosity of the particles and on the critical mass ratio. For example, Paper I finds fragmentation in collisions between a porous aggregate and a solid wall, whereas Langkowski et al. (2008) find sticking of a porous projectile by penetrating an also porous target.  Likewise, Hei\ss elmann et al., in prep. find bouncing of two similar-sized, porous dust aggregates, while Langkowski et al. (2008) find sticking for the same velocity where one collision partner (target) was significantly bigger. To address the importance of the mass ratio and porosity, we have identified eight different collision regimes (look-up tables) based on a binary treatment of porosity and mass ratio: i.e., (i) similarly sized or differently sized collision partners and (ii) porous or compact collision partners. The further distinction between target, which we always define as the heavier collision partner, and projectile then results in eight different collision regimes. We denote these regimes as {\pP} (porous projectile, porous target; target significantly bigger than the projectile), {\pc} (porous projectile, compact target; target of similar size than the projectile), etc.  

In Paper I we have classified each of these 19 experiments into one or more of these eight regimes (see Fig. 10 of Paper I). Based on extrapolation of experimental findings, we decide in which mass and velocity range collisions result in sticking, bouncing, or fragmentation. These results are presented in Fig. 11 in Paper I.

It should be noted that the critical mass ratio between the equal-size ({\pp}, {\cc}, etc.) and the different size regimes ({\pP}, {\cC}, etc.) is ill-constrained by experiments. Therefore, we use critical mass ratios of $r_m$ = 10, 100 and 1000 to explore the effect of this parameter.

\subsection{Porosity}

Paper I defined a binary representation of the porosity, particles are either porous or compact. Following the simple model of \cite{Weidling2009}, we include a continuous transition between these two `phases'. They showed that the compaction of particles due to bouncing can be described by porous and compacted sites on the surface of the aggregate. A site of the aggregate is porous if it did not encounter any collisions yet (e.g. bouncing), a compacted site encountered at least one collision already but any  further collision happening at that part of the surface cannot change the porosity of this site anymore. We describe the probability of hitting a passive site of the aggregate in the following way:
\begin{equation}
P_p = \frac{\phi_c - \phi}{\phi_c - \phi_p}, 
\end{equation}
where $\phi$ is the volume filling factor of the aggregate, $\phi_c$ is the critical porosity ($\phi_c=0.4$, see Paper I), and $\phi_p$ is the volume filling factor of the porous site, which is chosen to be 0.15. If $\phi$ is between 0.15 and 0.4, a random number decides whether the particle collided with a porous or a compact site. Such a treatment of the porosity ensures a continuous transition from porous to compact aggregates. \\

During the initial {\Sa} phase, particles are in the fractal growth regime (\cite{Ossenkopf:1993p80}, \cite{Blum2000a}, \cite{Krause2004}). Particles grow initially due to Brownian motion and later due to turbulence. The structure of the aggregate depends on whether the collision happened between a cluster and a monomer (PCA) or between two clusters (CCA). The latter results in fluffy aggregates with fractal dimension of 2, while the former leads to more compact structures with fractal dimension of 3. The hit\&stick recipe of \cite{Ormel:2007p93} attempts to interpolate between these two fractal models. At one point, collisional energies become large enough to invalidate this assumption. This occurs when the collision energy is five times higher than the rolling energy of monomers (\cite{Dominik:1997p89}, \cite{Blum2000}). The internal structure then becomes homogenous. In our model we assume that once the fractal growth due to S1 is over, {\Ba} restructures the aggregates producing compact structures with fractal dimension of 3. As our model can not follow the exact shape of the particles, we assume that the aggregates are spheres and can be described with a single density ($\rho$) thus neglecting the effects of e.g. the `toothing radius' of \cite{Ossenkopf:1993p80} or the craters forming during {\Sc}.

\subsection{The Monte Carlo method}

Using the expressions for the relative velocity, the collisional cross section between the dust particles, and the collisional outcome, we solve for the temporal evolution of the dust size distribution. Traditionally, the Smoluchowski equation is solved to follow the evolution of the mass distribution function (e.g. \cite{Dullemond:2004p325}, \cite{Dullemond:2005p78}, \cite{Tanaka2005}, \cite{Brauer2008a}). The continuous form of the Smoluchowski equation used in these works lacks the stochasticity of the coagulation problem (\cite{Safronov1969}). All bodies with mass $m$ will grow in the same way thus the spatial and temporal fluctuations of the particle ensemble are averaged out. In reality, however, particles with similar masses might follow a different evolutionary path depending on what other particles they collide with. The collision model typically used in these works is, by necessity, rather simple as in the Smoluchowski formulation the collision and time evolution steps are linked together. These collision models consist of sticking and fragmentation and only the mass of the particles is followed. The advantage of such a model is that it is computationally not too expensive: The entire disk can be modeled. 

\cite{Ormel:2007p93} introduced a new Monte Carlo method to solve for the mass and the porosity distribution function simultaneously. Their collision model consists of sticking and compaction; ZsD08 added a simple fragmentation model as well. Although these models are more detailed, one can see that they still lack the full complexity which is observed at ``the zoo'' of laboratory collision experiments.

The MC-approach used in this study has previously been presented by ZsD08. It can be characterized by two key properties: (1) the number of MC-particles (also referred to as representative particles) is kept constant; (2) the method follows the mass of the particle distribution.

Property (1) is required to preserve good statistics. Because of the $\sqrt{N}$ noise of MC-methods, a large fluctuation of $N$ would severely affect the accuracy of the method (\cite{Ormel:2008p95}). The second property states that our primary interest lies in the particles that contain most of the mass of the system.  Moreover, it has been shown that following the particle's mass distribution -- rather than the number distribution -- is also a prerequisite to preserve a good correspondence with systems that experience strong growth (\cite{Ormel:2008p95}).

Property (2) ensures that the MC method samples the parameter space only where a signiÞcant portion of the total dust mass is. However, this is not always desirable. For instance, radiative transfer calculations require the surface area distribution of the aggregates, which determines the opacity. If most of the particle mass is contained in big particles (which are not observable) the amount of small particles (which could contain most of the surface area and determines the IR appearance of the disk) might be resolved with a bad statistics. But if we are interested in following the evolution of the dominant portion of the dust, then MC methods naturally focus on these parts of the phase space. 

A required condition for the ZsD08 method to work is that the number of the representative particles $N$ is much less than the number of actual aggregates present in the system under consideration -- a condition that is safely met in any of our simulation runs. Then, a representative particle will collide only with the non-representative particles, whose distribution is assumed to be the same as that of the representative particles. We refer to ZsD08 for details regarding the precise implementation and accuracy of the method; here we further concentrate on how the method operates under the new collisional setup.\\

The collision kernel is defined as the product of the cross section of the colliding particles and their relative velocity:
\begin{equation}
K_{i,k}=\sigma_{i,k} \Delta v_{i,k},
\label{eq:kernel}
\end{equation}
where the index $i$ corresponds to the representative particle and $k$ is the index of the non-representative particle. The kernel is proportional to the probability of a collision. The value of $K_{i,k}$ is calculated for every possible particle pair, and random numbers determine which of the collision will occur first and at which time interval.\\

The above properties and conditions specify the essence of the ZsD08 method: one of the two collision particles is a representative particle and, by property (1), only one of the collisional products becomes the new representative particle. By property (2) the choice for the new representative particle is weighed by the mass of the collision products. A very helpful analogy here is that of the representative `atom', which is contained within the representative particle. The choice for the new representative particle after the collision is then proportional to the probability of the representative `atom' ending up in the collision products. If, for instance, a collision leads to the production of an entire distribution of debris particles, the probability that a particular debris fragment becomes the new representative particle is proportional to the likelihood of this fragment to contain the representative `atom'.

\subsection{Implementation of the collision types}
\label{sect:implement}

We describe the implementation of the collision model using the representative `atom' concept. We refer to Paper I for details of the various collision types described below. 

\paragraph{{\Sa}, {\Sb}, {\Sc}: }All three of these collision types result in sticking and increase the mass of the aggregate by that of the projectile, but the porosity changes in a different manner (see Paper I). The new mass of the representative particle $i$ is then the sum of the original particle masses, $m_{i, \mathrm{new}}=m_i+m_k$, where $m_i$ is the mass of the representative particle and $m_k$ is the mass of the non-representative particle. 

\paragraph{\Sd : } In the case of {\Sd}, a certain percentage of the mass of the projectile sticks to the target, while the left-over mass of the projectile will fragment into a power law distribution (see Paper I). 

There are two situations to consider:
\begin{enumerate} 
\item The representative `atom' is part of the target. The mass of the new aggregate will be the mass of the original aggregate plus the transferred mass from the non-representative particle ($m_{i, \mathrm{new}}=m_i+m_{\mathrm{trans}}$, where $m_{\mathrm{trans}}$ is the transferred mass calculated according to Paper I).
\item The representative `atom' is part of the projectile. Again, we have two situations.
\begin{enumerate} 
\item The representative `atom' will be transferred to the non-representative particle. The mass of the new representative particle will be the mass of the non-representative particle plus the transferred material ($m_{i, \mathrm{new}}=m_k+m_{\mathrm{trans}}$). The probability of transferring (removing) the representative atom from the projectile is simply $P=m_{\mathrm{trans}}/m_i$, the ratio between the transferred mass and the mass of the projectile.
\item The representative `atom' remains in one of the fragments. The probability of this event is $P=(m_i-m_{\mathrm{trans}})/m_i$, the ratio between the fragmented mass to the original mass of the representative particle. As discussed in Paper I, the fragments follow a power law mass distribution. The distribution is defined by the maximum mass of the fragments, which is a function of the relative velocity and the total mass of the fragments. The total mass of the fragments is $m_i-m_{\mathrm{trans}}$. We randomly choose from the fragment distribution to find the new mass of the representative particle (to find which of the fragments will contain the representative `atom'). 
\end{enumerate}
\end{enumerate}

\paragraph{{\Ba}: } Upon {\Ba} particles collide and bounce. Bouncing itself does not change the mass of the particles, but it compactifies them according to Paper I. As observed in laboratory experiments (\cite{Weidling2009}), there is a small probability ($P_{\mathrm{frag}}=10^{-4}$) that the bouncing particle will break apart. If this happens, we break the particle into two equal mass pieces.

\paragraph{{\Bb}: } {\Bb} is, from the implementation point of view, similar to {\Sd}. The recipe to define the new representative particle is as in {\Sd}. The difference is that the projectile does not fragment during the collision, and that the porosity changes differently (see Paper I).  

\paragraph{{\Fa}: } Fragmentation only happens between similar sized aggregates in the {\pp} and {\cc} regimes. The fragments follow a power law mass distribution where the maximum mass of the fragments is determined by the relative velocity of the particles and the total mass that goes into the fragments (Paper I). We randomly choose from these distribution to determine the new mass of the representative particle.

\paragraph{{\Fb}: } {\Fb} happens between different sized particles only. During the collision the projectile``kicks out" pieces from the target aggregate. These pieces follow a power law distribution (see Paper I). We have to consider two cases. 
\begin{enumerate} 
\item The representative `atom' is in the target. Again, we have two possibilities.
\begin{enumerate} 
\item The representative `atom' will stay in the target after the collision. The mass of the new particle will be $m_{i, \mathrm{new}}=m_i-m_{\mathrm{er}}$, where $m_{\mathrm{er}}$ is the eroded mass. The probability of this event is $P=(m_i-m_{\mathrm{er}})/m_i$, that is the ratio between the left-over mass (which does not erode) and the mass of the original particle.
\item The representative `atom' is part of the eroded particles. As the eroded particles follow a power law distribution, we randomly pick from this distribution to determine the new mass of the representative particle. The likelihood for this event is the ratio between the eroded mass and the original mass of the particle ($P=m_{\mathrm{er}}/m_i$). 
\end{enumerate} 
\item The representative `atom' is part of the small particle which caused the erosion. As the particles do not stick and the small particle does not fragment, the representative particle remains unaffected. 
\end{enumerate}

\paragraph{{\Fc}: } In {\Fc} the porous particle gets destroyed by the compact one and transfers a certain amount of mass to the compact particle. {\Fc} only happens in the {\cp} regime. Again, we have two possibilities. 
\begin{enumerate} 
\item The representative `atom' is part of the compact particle. In this case the representative `atom' cannot leave the particle. The new mass of the representative particle will be $m_{i, \mathrm{new}}=m_i+m_{\mathrm{trans}}$, the sum of the original mass plus the transferred mass. 
\item The representative `atom' was part of the porous aggregate. 
\begin{enumerate} 
\item The representative `atom' is part of the material which is transferred to the compact particle. In this case, the new mass of the particle will be that of the compact (non-representative) particle plus the transferred material ($m_{i, \mathrm{new}}=m_k+m_{\mathrm{trans}}$). The probability of this event is $P=m_{\mathrm{trans}}/m_i$. 
\item The representative `atom' is part of the fragments. As before, the mass distribution will follow a power law and we randomly pick from this distribution to determine the new mass of the representative particle. The probability of this event is $P=(m_i-m_{\mathrm{trans}})/m_i$. 
\end{enumerate} 
\end{enumerate}

\subsection{Evolving the particle properties in time}
We summarize how the particle properties are evolved in time using the above described kernel. We start with the size and porosity distribution of the particles at a given time, $t$. At $t=0$, we must give the initial size and porosity distribution, see Sect. \ref{sec:inicond}. Knowing these:
\begin{itemize}
\item We calculate the cross sections of all possible collision partners, as well as their relative velocities using the equations described in Sect. \ref{sec:vrel}. Both are used to determine the collision rates between the particle pairs. 
\item By using random numbers, we identify from the collision rates the representative particle, which is involved in the collision, as well as the non-representative particle it collides with and at what time the collision takes place ($t+\Delta t$).
\item Knowing the masses (mass ratio) and porosities of the collision partners, we identify in which of the eight regimes the collision takes place (e.g. {\pP}, or {\pC}, etc.). 
\item Next, we identify which of the nine collision types materializes (Fig. \ref{coltypes}) using the relative velocity of the particles and the mass of the projectile (see Paper I).
\item Based on the collision recipe described in Paper I and Sect. \ref{sect:implement}, the new mass and new porosity of the representative particle is calculated and the new size and porosity distribution of the particles at time $t+\Delta t$ is obtained.
\item In the final step, we update the collision rates.
\end{itemize}

\subsection{Numerical issues}

As mentioned in ZsD08, a sufficiently high number of representative particles is needed to properly reproduce the physics of the collision kernel. We performed simulations with an increasing number of representative particles and found that for more than 200-300 particles the results of the simulations do not change significantly. For all of the simulations described in the following sections 500 representative particles are used and we average the results of 20 simulations to decrease the numerical noise of the code.\\

The required computational time strongly depends on the collision rate of the particles thus determined by the dust density and the relative velocity (the $\alpha$ turbulence parameter mostly) and the length of the simulation. On a 2.83 GHz CPU, running the simulations on a single core, the CPU time varies between nine hours (the low density model with $\alpha = 10^{-5}$) and three days (the high density model with $\alpha=10^{-3}$). Both simulations covered $10^6$ years of particle evolution. The high density simulation with $\alpha = 10^{-4}$, critical mass ratio of 100 and $t=10^7$ years of evolution takes twelve days to simulate.

\section{Results}
\label{sec:res}
\subsection{Initial conditions, setup of simulations}
\label{sec:inicond}
All simulations start with silicate monomers of 1.5 $\mu$m diameter and 2 g cm$^{-3}$ material density (monodisperse size distribution). We simulate the dust evolution at the midplane of our disk models at a distance of 1 AU from the central star. The gas density is obtained from the disk models described in Sect. \ref{sec:disks}. We assume a typical 1:100 dust to gas ratio. We follow the history of each collision: the mass and porosity of the colliding particles, their relative velocity, the occurred collision type and the new mass and porosity of the particles. In this way we can reconstruct the history of the dust evolution.

The parameters we vary in this study are the gas density $\rho_g$ and the turbulence parameter $\alpha$. We also treat the critical mass ratio $r_m$ as a free parameter in order to explore its effect on the dust evolution.

We provide a detailed description of the low density model with $\alpha = 10^{-4}$ and critical mass ratio of 100 in Sect. \ref{sec:lowd}. We then compare this with the MMSN model and the high density model using the same turbulence parameter and the critical mass ratio (Sects. \ref{sec:mmsn} and \ref{sec:desch}). In Sects. \ref{sec:turb} and \ref{sec:critmass} we discuss the effects of changing the turbulence parameter and critical mass ratio by comparing those results with the two example runs.

\subsection{The low density model}
\label{sec:lowd}

\begin{figure*}
\centering
  \includegraphics[width=0.9\textwidth]{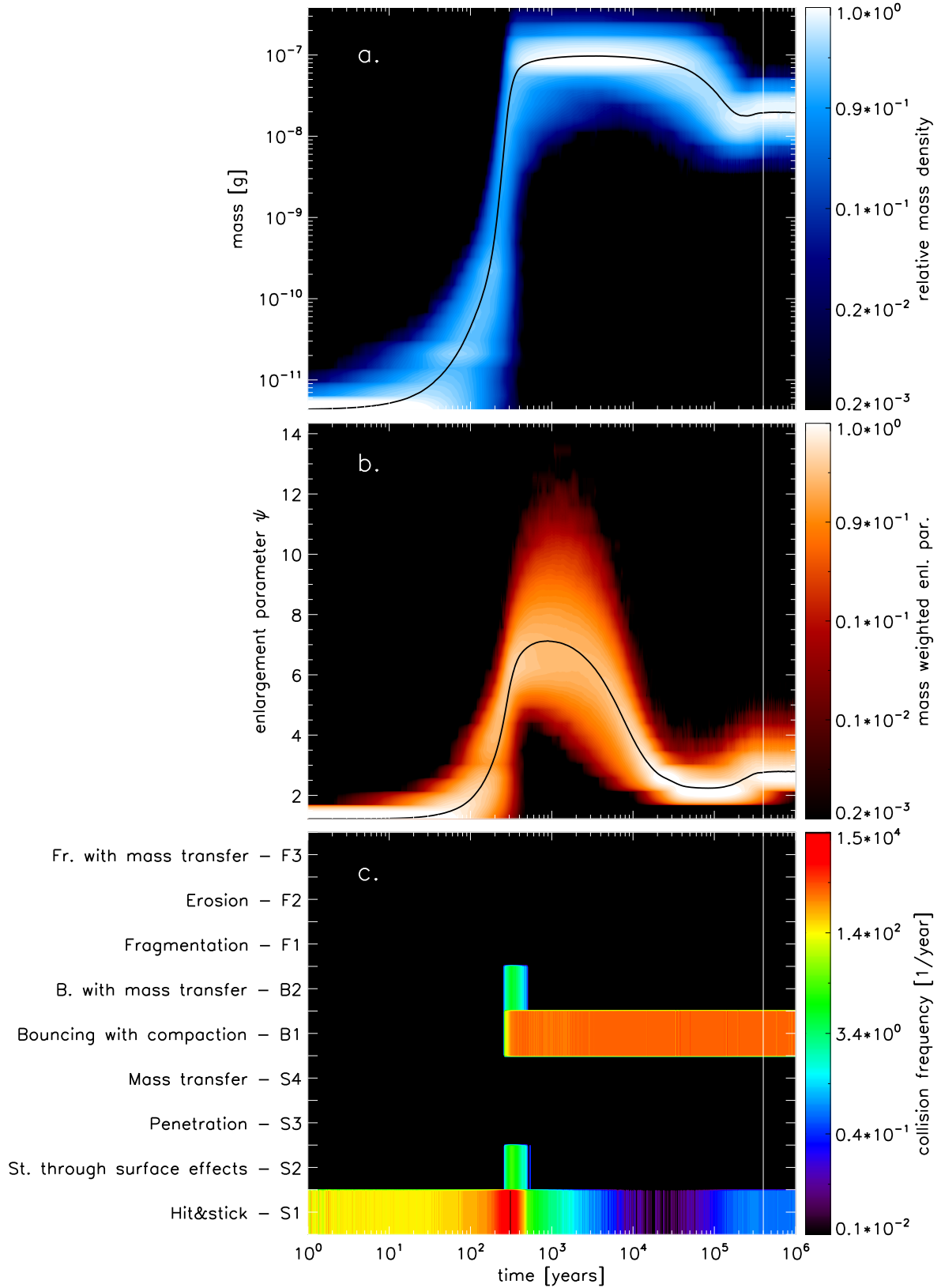}
  \caption{The evolution of the mass distribution (a), enlargement parameter distribution (b) and the collision frequency of the nine different collision types (c) in the low density model with $\alpha=10^{-4}$ and critical mass ratio of 100. The x-axis shows the time. The y-axis of the (a) and (b) figures show the logarithmic mass and the linear enlargement parameter respectively. The contours represent the normalized mass density and the mass weighted enlargement parameter. The black lines represents the average of the mass and enlargement parameter at a given time. The y-axis on the (c) figure represents the nine collision types. Each stripe shows the total collision rate of the collision types. Two distinct phases can be distinguished. During the initial 300 yr particles grow by {\Sa}, after that the evolution is governed by {\Ba}. The white lines indicate how long our `local approach' assumption is valid (discussed in Sect. \ref{sec:impl}).}
  \label{lowd_pics}
\end{figure*}

\begin{figure*}
\centering
  \includegraphics[width=0.9\textwidth]{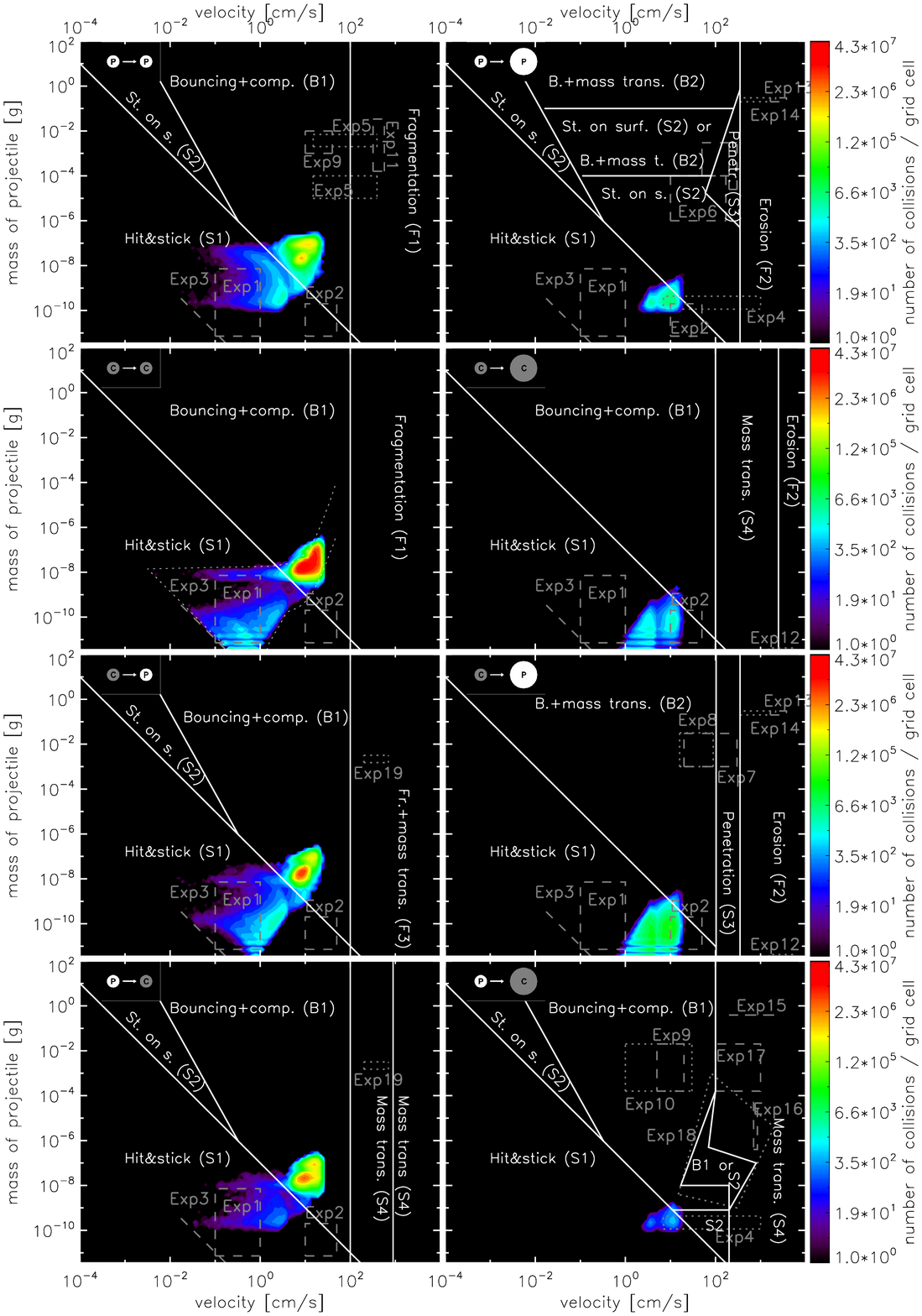}
  \caption{The collision history of the eight regimes in the low density model for $\alpha=10^{-4}$. The x-axis is the relative velocity, the y-axis shows the projectile mass. The different collision types, their border lines, as well as the areas covered with laboratory experiments (grey) are plotted. A relative velocity - mass grid is created and in these grid cells we calculate how many collisions happened until the `local approach' assumption is valid ($4\times 10^5$ yr). This is represented by the colors: yellow and red indicate a high collision frequency. The two dotted lines on the {\cc} regime are evolution tracks. Assuming a constant (40\%) volume filling factor, the relative velocity between equal sized particles (left curve) and particles with a mass ratio of 100 (right curve) can be calculated. The collisions in the simulation should lay between these two lines. The small deviations are due to the fact that the volume filling factor is not exactly 40\% during the simulation.}
  \label{lowd_cont}
\end{figure*}

The gas density in this disk model at 1 AU is $2.4 \times 10^{-11}$ g cm$^{-3}$, the turbulence parameter is $\alpha=10^{-4}$, the critical mass ratio is $r_{\mathrm{m}}=100$. As shown in Fig. \ref{relcont}, the particles reach the fragmentation velocity (1 m s$^{-1}$) already at sizes smaller than millimeter because the particles in low gas density environment decouple from the gas already at these small radii.

Figure \ref{lowd_pics} shows the evolution of the mass distribution (a), the porosity distribution (b) and the collision frequency of the various collision types (c). The x-axis shows the time in a logarithmic scale. The y-axis of Fig. \ref{lowd_pics}a, b shows the mass and enlargement distributions, respectively. Here, the intensity of the color reflects the number density of representative particles, which, as explained in ZsD08, measures the mass density of the distribution. Thus, in Fig. \ref{lowd_pics}a the intensity levels directly reflect the mass density, while in Fig. \ref{lowd_pics}b the colours indicate the mass weighted enlargement parameter. The black lines show the average of these quantities over the particle distribution. The y-axis in Fig. \ref{lowd_pics}c represents the nine collision types used in this paper. Every stripe shows the total collision rate of the collision types at a given time.

Figure \ref{lowd_cont} represents the collision history in the eight collision regimes. The x-axis is the velocity, the y-axis shows the mass of the projectile. A mass-velocity grid is created and for all grid cells we calculate how many collisions happened inside that given grid cell during which our `local approach' assumption is correct, that is $4\times 10^5$ yr. The different collision types and their border lines, as well as the areas which are covered with laboratory experiments (indicated with grey colors) are plotted. For more details on the experiments, see Paper I. 

In the {\cc} panel, we indicate two curves with dotted lines. These curves are evolution tracks. The left curve is obtained by calculating the relative velocity between equal sized particles with an enlargement parameter of 2.5 (volume filling factor of 40\%). The right curve represents the relative velocity between particles having a mass ratio of 100. These two curves serve as a guide to our results, as collisions should happen between these two curves in the {\cc} panel. The lower part of the left curve, where the relative velocity decreases with increasing mass, is a sign that relative velocities between equal sized particles are dominated by Brownian motion. For higher masses, the relative velocity is dominated by turbulence. These curves do not precisely match the contours because we assumed a constant enlargement parameter of 40\% when calculating the evolution tracks, whereas $\Psi$ is a free parameter in the simulation.

%It is also notable that there are no collisions for particle masses less than 10$^{-10}$ g in the {\pp} and {\pP} regimes. The enlargement parameter as a function of monomer number is an increasing function. One single monomer has an enlargement parameter of 1 (100\% volume filling factor), an aggregate consisting of 100 monomers has an enlargement parameter of roughly 3 (33\% volume filling factor). Therefore, collisions between aggregates consisting low number of monomers appear on the other six regimes. The very first collisions occur in the {\cc} regime. The intermediate stages of the Hit\&Stick phase are happening on the {\pp} and {\pP} regimes. When particles enter the second phase governed by {\Ba}, the collisions are again happening mostly on the {\cc} regime, as the enlargement parameter of the particles is low at the end of the simulation.\\

\subsubsection{Early evolution}
We discuss here the evolution of the distribution functions until the `local approach' assumption becomes invalid ($4\times 10^5$ yr). The long term evolution of the dust is discussed in Section \ref{subsec:longt}.

We distinguish two distinct phases here. During the first 300 yr, particles grow by the {\Sa} mechanism. The second phase is {\Ba} dominated; the particles leave the S1 regimes. During this phase the mass of the particles is slowly decreasing and the enlargement parameter asymptotically reaches a minimum value of 2.23. As discussed in Paper I, keeping the bouncing velocity of a particle constant, the porosity of the aggregate will asymptotically reach a maximum value, $\phi_{\mathrm{max}}$ (see Paper I). The relative velocity of a particle is a function of the friction time (Eq. \ref{eq:ts1}), which depends on the ratio of the mass to surface area, $m/A$. Since particle growth is halted at this point in the simulation ($m$ stays constant), only a decrease in $A$ due to compaction can further increase the velocity between particles. The particle radius can decrease until either $\phi_{\mathrm{max}}$ for the given relative velocity is reached, or until particles reach the maximum compaction possible. The latter limit, random close packing (RCP), corresponds to an enlargement parameter of 1.6 (volume filling factor of $\sim$60\%).

We find that fragmentation does not play a role during the evolution of these particles indicated by Fig. \ref{lowd_pics}c. As can be seen in Fig. \ref{lowd_cont}, their evolution is halted by bouncing before the particles could reach the fragmentation barrier. The two dominant collision types are {\Sa} and {\Ba}.\\

\subsubsection{Termination of growth}
As we can see from Fig. \ref{lowd_cont}, sticking at higher energies than the {\Sa} border lines is only possible inside the {\pP} regime. As soon as we no longer have collisions inside this regime or the S1 regimes, the growth is halted. There can be two reasons why this is happening: 1.) All particles are compact; there are simply no collisions in the {\pP} regime. 2.) The width of the particle mass distribution is less than the critical mass ratio ($r_m$), such that all collision take place in the equal-size regimes ({\pp}, {\pc}, etc.).

In the case of the current simulation, the small particles have been `consumed'. Once the heavy particles grow into the {\Ba} area of the {\pp} regime, their growth in the {\pp} regime stops. The heavy particles collect the small ones via collisions in the {\pP} regime and by doing so, the width of the distribution is reduced to a value which is less than $r_m$. Therefore, before particles could reach the fragmentation barrier, growth is halted. Due to B1, particles get compacted and collisions in the {\cc}, {\cp} and {\pc} regimes appear.

\subsubsection{Long term evolution}
\label{subsec:longt}
Before discussing the long term evolution of the distribution functions, we must consider for how long our starting assumptions (`local approach' and constant gas density) hold true.

Using Eq. \ref{eq:drift}, we calculate that a particle with Stokes number $10^{-4}$ drifts a distance of 1 AU in roughly $4\times 10^5$ yr. This is the drift timescale beyond which the `local approach' assumption (discussed in Sect. \ref{sec:impl}) is not valid anymore: particles become separated from each other on this timescale.

Another process through which particles separate is by viscous spreading. We determine the viscous timescale of the disk at 1 AU:
\begin{equation}
t_{\mathrm{vis}}=r^2/ \nu_T,
\end{equation}
where $r$ is the distance from the central star (1 AU), $\nu_T$ is defined in Eq. \ref{eq:nuT}. The viscous timescale in our model, using $\alpha = 10^{-4}$, is of the order of $10^6$ yr. 

One has to consider the results of the simulation with caution for longer times than the drift or viscous timescales. The equilibrium or final state of the particles is reached when mass decrease during {\Ba} and {\Bb} and mass increase by {\Sa} and {\Sb} are in equilibrium. Or in other words, the final state is reached when the evolution of the average mass and the enlargement parameter is only determined by the stochastic fluctuations of the simulation. We find that the equilibrium state of the particles is hardly reached within these timescales. Upon neglecting these warnings, we find that the equilibrium state of the dust is reached at $t=4 \times 10^5$ yr. The equilibrium is reached between the bouncing collisions resulting in breakage and {\Sa} (see Fig. \ref{lowd_pics}c). The equilibrium average mass and porosity of the particles are $ \bar{m}_{\mathrm{fin}}=2\times 10^{-8} \mbox{ g}$, $\bar{\Psi}_{\mathrm{fin}}=2.77$.\\

To be able to compare the distribution functions of different runs, we define some quantities using the mean of the distribution functions shown with black lines in Fig. \ref{lowd_pics}a and b: $\max (\bar{m})$, the maximum of the mean mass; $\max (\bar{\Psi})$, maximum of the mean enlargement parameter; $\Psi _{min}$, the minimum mean enlargement parameter when particles do not compact anymore; $t_{\mathrm{noc}}$, the time when $\Psi _{min}$ is reached, that is when  the time derivative of $\bar{\Psi}$ is zero ($d\bar{\Psi}/dt = 0$); and $\max(\bar{St})$, the maximum average Stokes number reached during the simulation. The values of these quantities are listed in Table \ref{tab:res} (model id `Lt1d-4m100'). In this table, Col. 1 describes the model names. `L' stands for the low density model, `M' is the MMSN model, `H' is the high density model, the letter `t' and the following number indicates the value of the turbulence parameter, and the letter `m' and the number shows the used critical mass ratio values. Columns 2, 3 and 4 show the gas density, turbulence parameter and the critical mass ratio respectively. Columns 5, 6, 7, 8 and 9 list the parameters defined to characterize the distribution functions. These are $\max (\bar{m})$ in Col. 5, $\max (\bar{\Psi})$ in Col. 6, $\Psi _{min}$ in Col. 7, $t_{\mathrm{noc}}$ in Col. 8 and finally $\max(\bar{St})$ in Col. 9. 

\subsection{The MMSN model}
\label{sec:mmsn}

\begin{figure*}
\centering
  \includegraphics[width=0.9\textwidth]{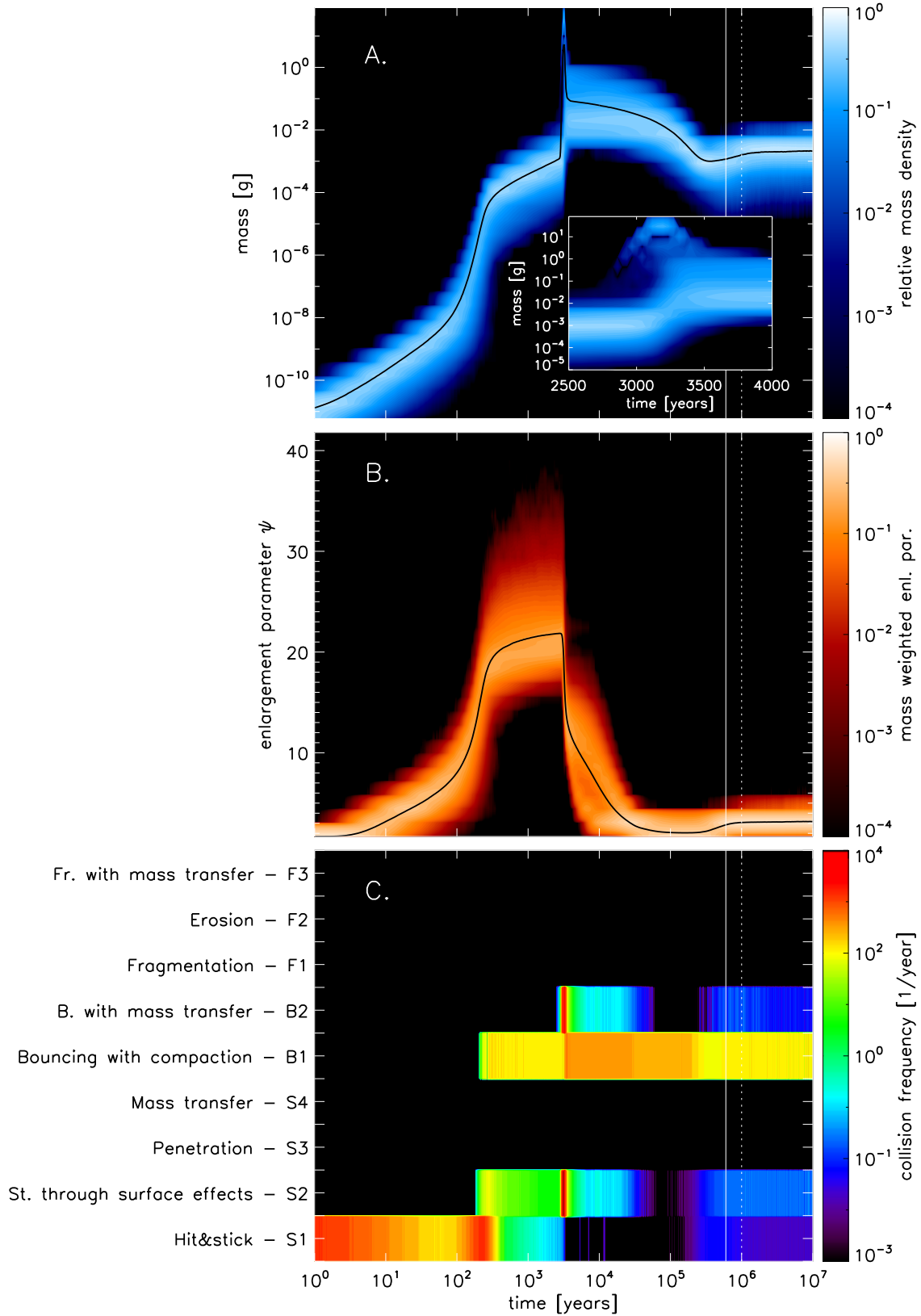}
  \caption{Same as Fig. \ref{lowd_pics} but for the MMSN model. We magnify the spike of the mass distribution at $\sim 2.5\times 10^{3}$ yr in Fig. (a). Four phases can be distinguished here. Initially (first 300 yr) particles grow purely by {\Sa}. After this the growth slows down because Bouncing with compation (B1) starts and all particles leave the {\Sa} regime. Between $3\times 10^3$ and $10^4$ yr, particles enter the transition regime between {\Sb} and {\Bb} on the {\pP} regime. Some particles reach masses of 1 g, but their masses are fastly reduced by B2. The last phase is {\Ba} dominated. The solid/dotted white lines indicate how long our `local approach' assumptions are valid (discussed in Sect. \ref{sec:impl}).}
  \label{mmsn_pics}
\end{figure*}

\begin{figure*}
\centering
  \includegraphics[width=0.9\textwidth]{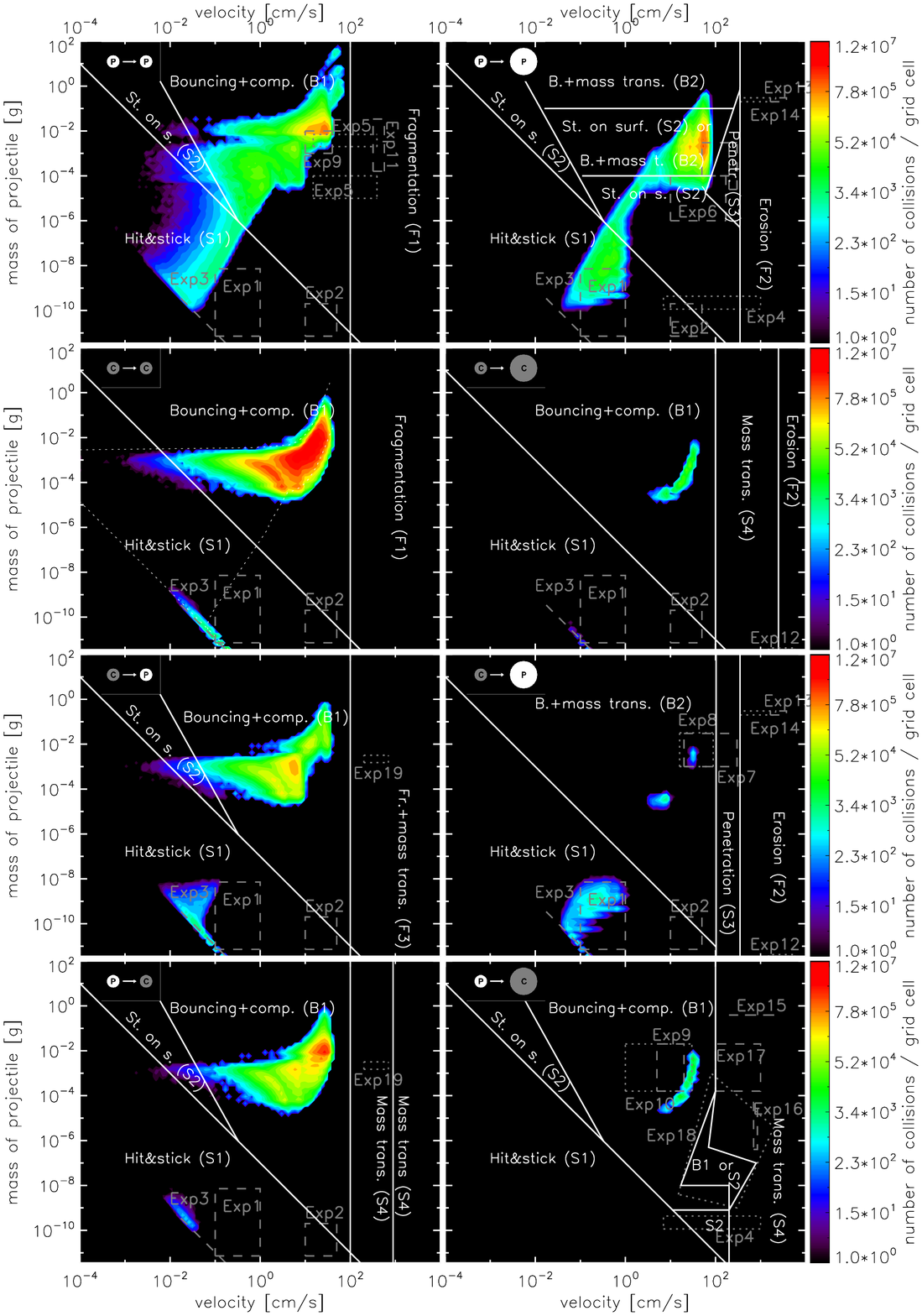}
  \caption{Same as Fig. \ref{lowd_cont} but for the MMSN model. The particles are better coupled to the gas due to the higher gas density. Therefore, they grow to bigger sizes than in the low density model.}
  \label{mmsn_cont}
\end{figure*}

The gas density in the MMSN model at 1 AU at the midplane is $1.4\times 10^{-9}$ g cm$^{-3}$, $\alpha = 10^{-4}$, the critical mass ratio is 100. As shown in Fig. \ref{relcont}, the particles grow to bigger sizes than in the low density model, as they are better coupled to the gas and the relative velocities are suppressed. As in the previous Section, we first discuss the evolution of the distribution functions for as long as the `local approach' assumption holds true ($6\times 10^5$ yr in this model).

Figure \ref{mmsn_pics} shows again the time evolution of the mass (a), enlargement parameter (b), and the collision frequency (c). Figure \ref{mmsn_cont} shows the collision history. These figures show a rather different evolution than the previous model.

\subsubsection{Early evolution}
\label{subsec:mmsn_early}
We find that during the fractal growth regime, the collision rate of {\Sa} is much higher than in the low density model (Fig. \ref{mmsn_pics}c). This is due to the higher dust densities. We can see from Fig. \ref{mmsn_cont}, {\cc} regime, that growth starts with Brownian motion because the relative velocity decreases with increasing particle mass for particle masses less than $10^{-9}$ g. As a result of these low velocity collisions, some particles reach enlargement parameter values higher than 30 (volume filling factor less than 3.3\%). At 200 yr, some particles grow above the border line of {\Sa} and enter the area of {\Sb} in the {\pP} plot, and {\Ba} in the {\pp} plot. Growth due to S1 and S2 continues until different sized particles enter the transition regime in the {\pP} plot. One can see in Fig. \ref{mmsn_pics}a, that some particles reach 1 g in mass. However when particle collisions enter the transition regime between {\Bb} and {\Sb} in the {\pP} plot, their masses are equalized due to the mass transfer of the B2 collisions and the collisions shift to the similar sized regime (B1). We find that after roughly $10^4$ yr particles mostly bounce and compact. The enlargement parameter reaches a minimum value of 1.85 (54\% volume filling factor), the mass distribution function slowly decreases due to a small probability of breakage. Collisions at this point are mainly happening in the {\cc} regime. \\

A peculiar feature of Fig. \ref{mmsn_pics}a is a peak at $t=2.5\times 10^3$ yr, which is accompanied by a fast decrease in the enlargement parameter in Fig. \ref{mmsn_pics}b and an increased collision rate of {\Sb} and {\Bb} in Fig. \ref{mmsn_pics}c. At this point, the relative velocity due to turbulence increases. As discussed in Sect. \ref{sec:vrel}, particles leave the `tightly coupled particle' regime and enter the `intermediate particle' regime (see the relative velocity bump in Fig. \ref{relcont}b). We calculate the growth timescale of the heaviest particle with mass $M$ in the simulation as follows:
\begin{equation}
t_{\mathrm{gr}}=\left( \frac{1}{M}\frac{dM}{dt} \right)^{-1}.
\end{equation}
This is illustrated with a dotted line in Fig. \ref{fig:acc_mmsn}. As a comparison, we also calculate the minimum growth timescale that a particle can have (solid line). That is:
\begin{equation}
t_{\mathrm{max}}= \left( \frac{M}{\rho_d \Delta v \sigma_M}\right)^{-1},
\end{equation}
where $\sigma_M$ is the cross section of the largest particle. Here, we assume that the `swept up' particles have masses of $M/100$, therefore we use the relative velocity curve presented in Fig. \ref{relcont}b, dotted line. The effect of the relative velocity bump and the increased growth rate is seen at 0.1 g. 

The relative velocity `boost' happens shortly after the particles enter the transition regime of S2 and B2 in the {\pP} plot. The heaviest particle, which encounters the velocity transition the earliest, experiences higher relative velocities leading to an increased collision rate with the other particles. As the particles are initially located at the lower part of the S2-B2 transition regime (with masses of $10^{-3}$ g, see Fig. \ref{mmsn_pics}a), the heaviest particle experiences fast growth and reaches masses of 30 g. The simulated timescale, however, does not reach the minimum growth timescale due to the {\Bb} collisions which are reducing the mass of the heaviest particle. The rest of the particle population increases in mass because of B2 and the growth rate of the heaviest particle decreases. Eventually, the fast growth of the heaviest particle is halted, the growth timescale at $m=30$ g is infinity. From this point on, the heaviest particle reduced in mass, and B2 equalizes the masses of the particles.\\

\begin{figure}
\centering
  \includegraphics[width=0.5\textwidth]{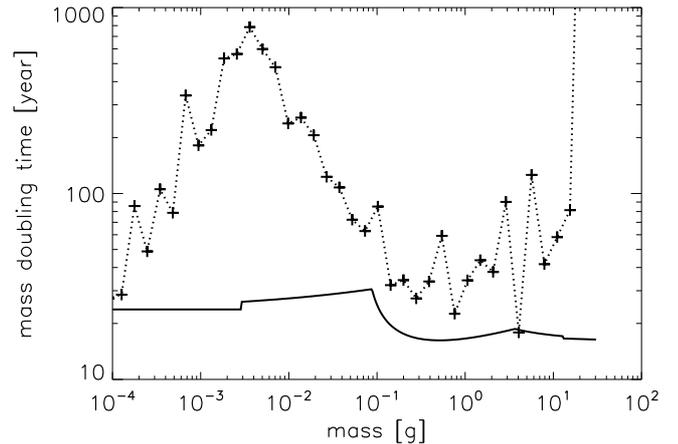}
  \caption{The dotted line and the `+' signs represent the growth timescale of the heaviest particle in the MMSN simulation with $\alpha=10^{-4}$ and $r_m = 100$. As a comparison, we show the minimum growth timescale a particle can have in this simulation (solid line).}
  \label{fig:acc_mmsn}
\end{figure}

\subsubsection{Long term evolution}
We calculate the drift and viscous timescales to determine how long our assumptions of `local approach' and constant gas density are valid. Assuming Stokes number $10^{-4}$ particles, we find that the drift timescale is of the order of $6\times 10^5$ yr, the viscous timescale is $10^6$ yr. These timescales are indicated with solid and dotted white lines in Fig. \ref{mmsn_pics}.

We find that the final equilibrium is reached at $t=2\times 10^6$ yr, which is longer than the drift and the viscous timescales. The equilibrium is reached between the growth mechanisms of {\Sa}, {\Sb} and the destruction mechanisms of bouncing resulting in breakage and {\Bb}. The final average mass and porosity of the particles are $\bar{m}_{\mathrm{fin}}=2\times 10^{-3} \mbox{ g}$, $\bar{\Psi}_{\mathrm{fin}}=3.3$.\\

We conclude that the dust evolution is more complex in the MMSN model than in the low density model because the complex interaction of the velocity field and the collision kernel is apparent in this model. As in the previous model, {\Ba} is the most frequent collision type and {\Sa} determines the initial particle growth, but {\Sb} and {\Bb} are of importance in this model. The final equilibrium is not reached within the drift and viscous timescales. 

\subsection{The high density model}
\label{sec:desch}
The gas density in this model is $2.7\times 10^{-8}$ g cm$^{-3}$ at the midplane of the disk at 1 AU distance from the central star. The values of $\alpha$, $r_m$ and the dust to gas ratio are the same as in the previous models.

Figure \ref{relcont}, dashed line, shows the relative velocity field of fluffy aggregates in this model. As already discussed in Sect. \ref{sec:vrel}, the aggregates reach 1 m s$^{-1}$ relative velocities at similar masses as the MMSN model due to the Stokes drag. Therefore, we expect that the final aggregate sizes and masses will be similar to the particles produced in the MMSN model.

Figure \ref{desch_pics} shows the time evolution of the mass (a), enlargement parameter (b) and the collision frequency (c). Figure \ref{desch_cont} illustrates the collision history. 

\subsubsection{Early evolution}
As seen in Fig. \ref{desch_cont}, Brownian motion is the dominant source of relative velocity, as long as particles stay below masses of $10^{-8}$ g (that is an order of magnitude higher than in the MMSN model). Therefore, the enlargement parameter of the aggregates is also higher than in the MMSN model. As the {\Sa} collisions are more frequent than in the MMSN model due to the higher dust densities, the particles reach the {\Sb} -- {\Bb} transition regime earlier, at $t=200$ yr. The peak in the mass distribution is not as pronounced as in the MMSN model. The relative velocity boost happens for heavier aggregates ($10^{-2}$ g, see Fig. \ref{relcont}b) due to the higher gas density of the model. When the fast growth of the heaviest particle starts, most of the projectiles are already in the transition regime. Here, the B2 collisions soon reduce the mass of the heaviest particle and narrow the mass distribution. 

In contrast to the MMSN model, the mass of the particles is not reduced due to the low probability of breakage in {\Ba}, but is kept nearly constant in time. This is the result of the increased collision rate of {\Sb}. The S2 collision rate increased because of low velocity collisions, which are occurring when particles are in the tightly couple regime and have similar stopping times. These S2 collisions are happening in the {\pp} regime as seen in Fig. \ref{desch_cont}. These collisions cancel out the effect of breakage in B1. 

The maximum Stokes number reached in this model is $3.6\times 10^{-5}$ (see Table \ref{tab:res}, model id `Ht1d-4m100'), lower than in the MMSN model. The growth in this model is halted by the {\Bb} collisions in the transition regime of the {\pP} panel. This shows us that particles cannot reach masses much larger than 1 g independently from the gas density (or Stokes number), because at this point, particles enter the S2-B2 transition regime and the growth is halted. Further increasing the gas density would result in even lower Stokes numbers. 

\subsubsection{Long term evolution} The drift and the viscous timescales in the high density model are both $10^6$ yr. As seen in Fig. \ref{desch_pics}a, the particle masses do not change significantly after $t=10^3$ yr. The porosity is reduced due to {\Ba} and it reaches a final value of 5.41 at $t=3\times 10^6$ yr.

\begin{figure*}
\centering
  \includegraphics[width=0.9\textwidth]{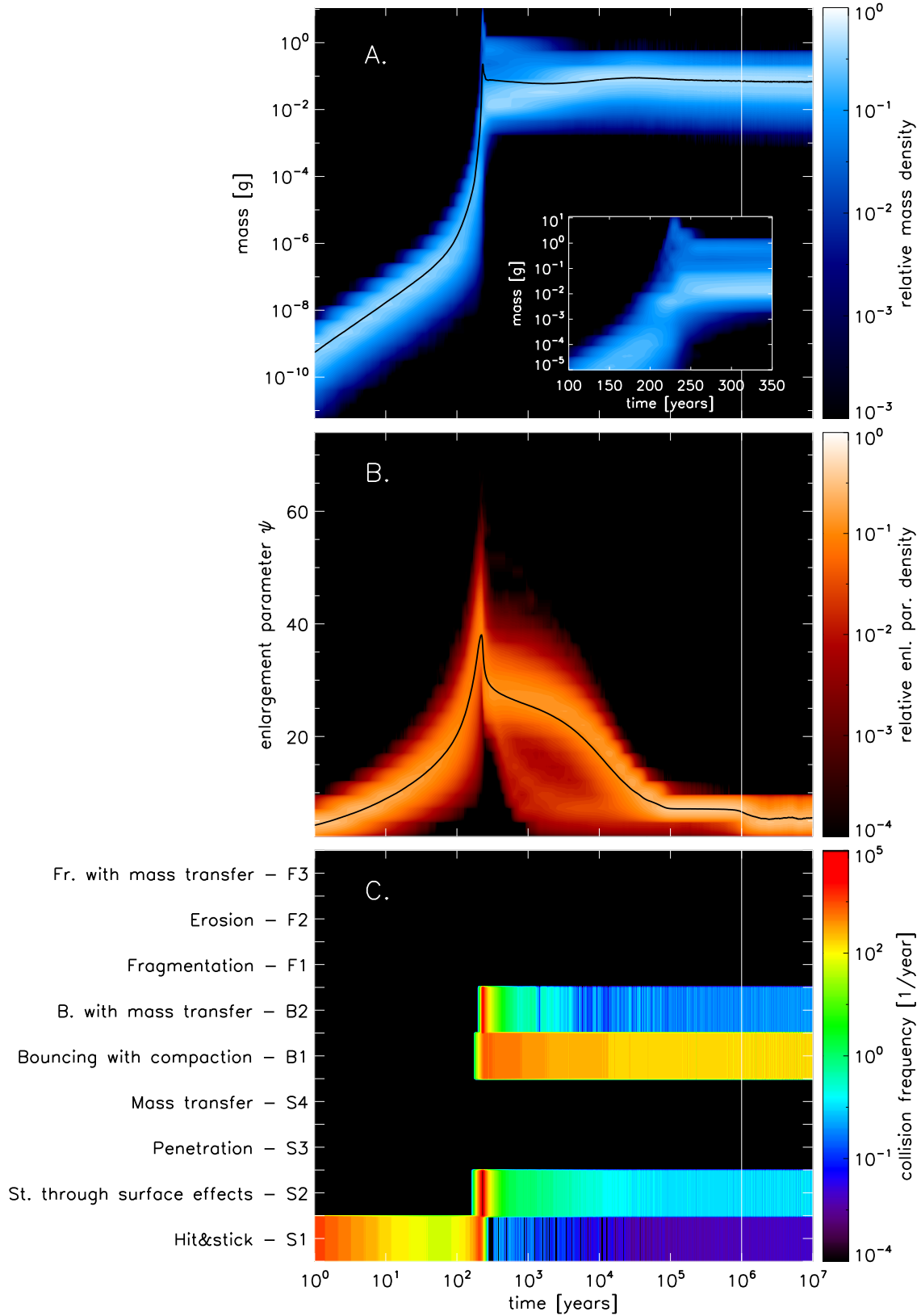}
  \caption{Same as Fig. \ref{lowd_pics} but for the high density model. As in Fig. \ref{mmsn_pics}a, we zoom in on the peak at the mass distribution. The solid white line indicate how long our `local approach' assumptions are valid at $t=10^6$ yr (discussed in Sect. \ref{sec:impl}).}
  \label{desch_pics}
\end{figure*}

\begin{figure*}
\centering
  \includegraphics[width=0.9\textwidth]{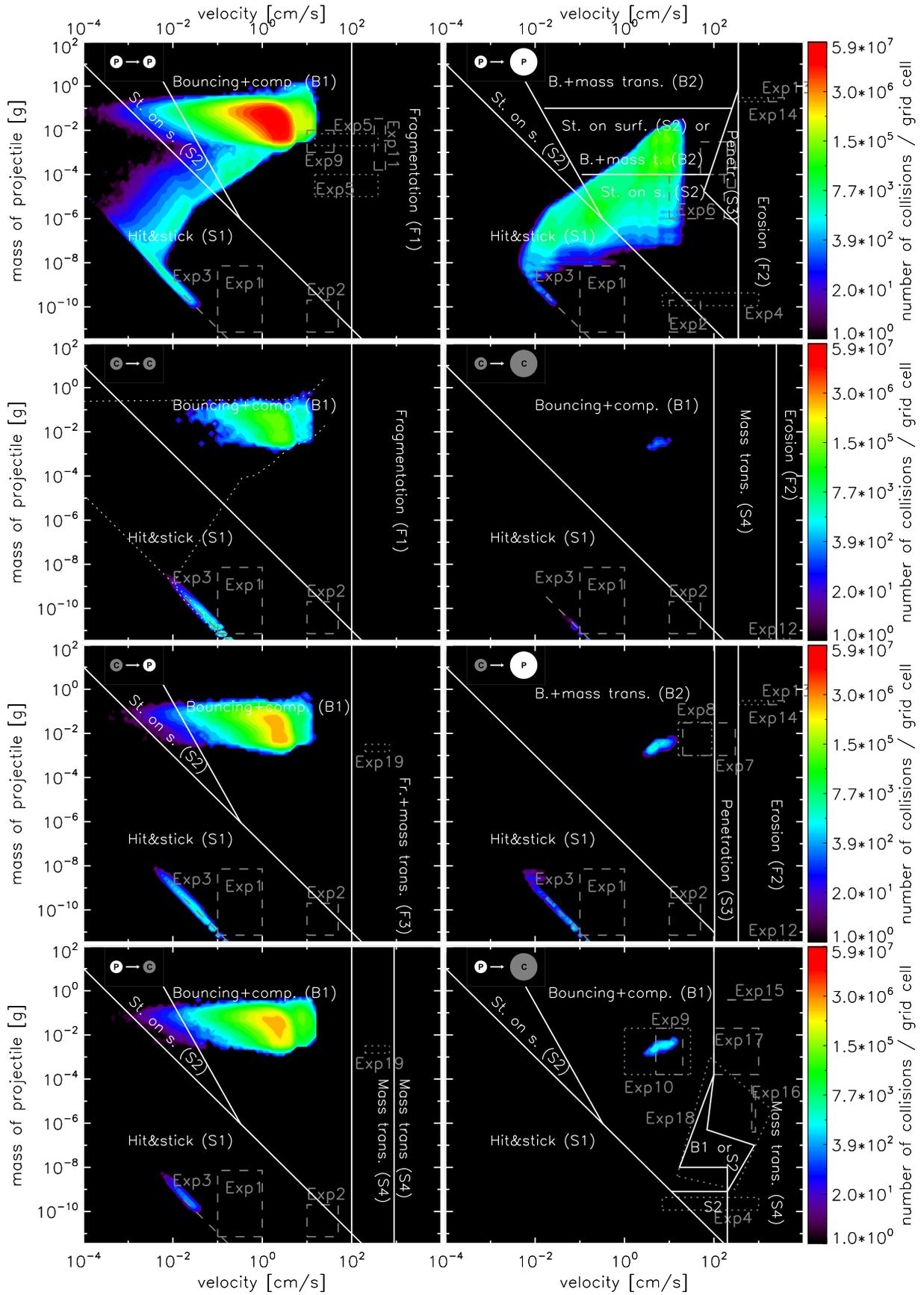}
  \caption{Same as Fig. \ref{lowd_cont} but for the high gas density model.}
  \label{desch_cont}
\end{figure*}

\subsection{Varying the turbulence parameter}
\label{sec:turb}

\begin{table*}
\caption{Overview and results of all the simulations.}
\label{table:lowd}      
\begin{center}   
\begin{tabular}{l l l l l l l l l }        
\hline\hline                 
Model& $\rho_g$ & $\alpha$ & $r_{\mathrm{m}}$	&$\max (\bar{m})$	&$\max (\bar{\Psi})$	&$\Psi _{\mathrm{min}}$	&$t_{\mathrm{noc}}$ & $\max (\bar{St})$\\    
	& [g cm$^{-3}$]& & &[g]	&	&	&[yr]	&\\
(1)	&(2)&(3)&(4)&(5)&(6)&(7)&(8)&(9)\\
\hline                        
Lt1d-3m100	&$2.4 \times 10^{-11}$	&$10^{-3}$	&100	&$8\times 10^{-8}$		&7.27	&1.77	&$2\times 10^{4}$	&$2.5\times 10^{-4}$	\\
Lt1d-4m100	&$2.4 \times 10^{-11}$	&$10^{-4}$	&100	&$9.7 \times 10^{-8}$	&7.12	&2.23	&$8\times 10^{4}$	&$2.2 \times 10^{-4}$	\\
Lt1d-5m100	&$2.4 \times 10^{-11}$	&$10^{-5}$	&100	&$2.66 \times 10^{-7}$	&7.72	&3.78	&$3 \times 10^{5}$	&$2.1 \times 10^{-4}$\\
\hline  
Mt1d-3m100	&$1.4\times 10^{-9}$	&$10^{-3}$	&100	&8.13				&24.41	&3.88	&$10^{4}$		&$5.1\times 10^{-4}$\\		
Mt1d-4m100	&$1.4\times 10^{-9}$	&$10^{-4}$	&100	&4.18				&21.9	&1.85	&$2\times 10^5$	&$2.8 \times 10^{-4}$\\		
Mt1d-5m100	&$1.4\times 10^{-9}$	&$10^{-5}$	&100	&$7.7\times 10^{-2}$	&30.0	&4.13	&$7\times 10^{5}$	&$2.1\times 10^{-4}$\\
\hline
Ht1d-3m100	&$2.7\times 10^{-8}$	&$10^{-3}$	&100	&3.77				&34.1	&5.61	&$10^5$			&$1.4 \times 10^{4}$\\		
Ht1d-4m100	&$2.7\times 10^{-8}$	&$10^{-4}$	&100	&0.23				&38.0	&5.41	&$3\times 10^6$	&$3.6\times 10^{-5}$\\		
Ht1d-5m100	&$2.7\times 10^{-8}$	&$10^{-5}$	&100	&0.28				&43.9	&4.94	&$4\times 10^6$	&$7.7\times 10^{-5}$\\
\hline
Lt1d-4m10	&$2.4 \times 10^{-11}$	&$10^{-4}$	&10		&$9.2\times 10^{-4}$	&5.88	&2.28	&$10^5$			&$3.8 \times 10^{-3}$	\\
Lt1d-4m100	&$2.4 \times 10^{-11}$	&$10^{-4}$	&100	&$9.7 \times 10^{-8}$	&7.12	&2.23	&$8\times 10^{4}$	&$2.2 \times 10^{-4}$	\\
Lt1d-4m1000	&$2.4 \times 10^{-11}$	&$10^{-4}$	&1000	&$9.7 \times 10^{-8}$	&7.09	&2.29	&$8\times 10^{4}$	&$2.2 \times 10^{-4}$	\\
\hline
Mt1d-4m10	&$1.4\times 10^{-9}$	&$10^{-4}$	&10		&$2.5\times 10^{-2}$	&19.4	&2.1		&$2\times 10^{5}$	&$2.2 \times 10^{-4}$\\
Mt1d-4m100	&$1.4\times 10^{-9}$	&$10^{-4}$	&100	&4.18				&21.9	&1.85	&$2\times 10^5$	&$2.8 \times 10^{-4}$\\		
Mt1d-4m1000	&$1.4\times 10^{-9}$	&$10^{-4}$	&1000	&$9.5 \times 10^{-3}$	&23.1	&2.9		&$2\times 10^{5}$	&$1.3\times 10^{-4}$\\
\hline  
Ht1d-4m10	&$2.7\times 10^{-8}$	&$10^{-4}$	&10		&0.15				&34.6	&2.46	&$2\times 10^6$	&$4.5\times 10^{-5}$\\		
Ht1d-4m100	&$2.7\times 10^{-8}$	&$10^{-4}$	&100	&0.23				&38.0	&5.41	&$3\times 10^6$	&$3.6\times 10^{-5}$\\		
Ht1d-4m1000	&$2.7\times 10^{-8}$	&$10^{-4}$	&1000	&$8.8 \times 10^{-2}$	&40.0	&7.1		&$10^5$			&$3.5\times 10^{-5}$\\
\hline
\label{tab:res}
\end{tabular}
\end{center}
In this table, Col. 1 describes the model names. `L' stands for the low density model, `M' is the MMSN model, `H' is the high density model, the letter `t' and the following number indicates the value of the turbulence parameter, the letter `m' and the number shows the used critical mass ratio values. Columns 2, 3 and 4 shows the gas density, turbulence parameter and the critical mass ratio respectively. Columns 5, 6, 7, 8 and 9 list the parameters defined to characterize the distribution functions. These are the average maximum mass in Col. 5, the average maximum enlargement parameter in Col. 6, the minimum enlargement parameter in Col. 7, the end of the compaction phase in Col. 8 and finally the average maximum Stokes number in Col. 9. 
\end{table*}

\begin{figure*}
\centering
  \includegraphics[width=0.9\textwidth]{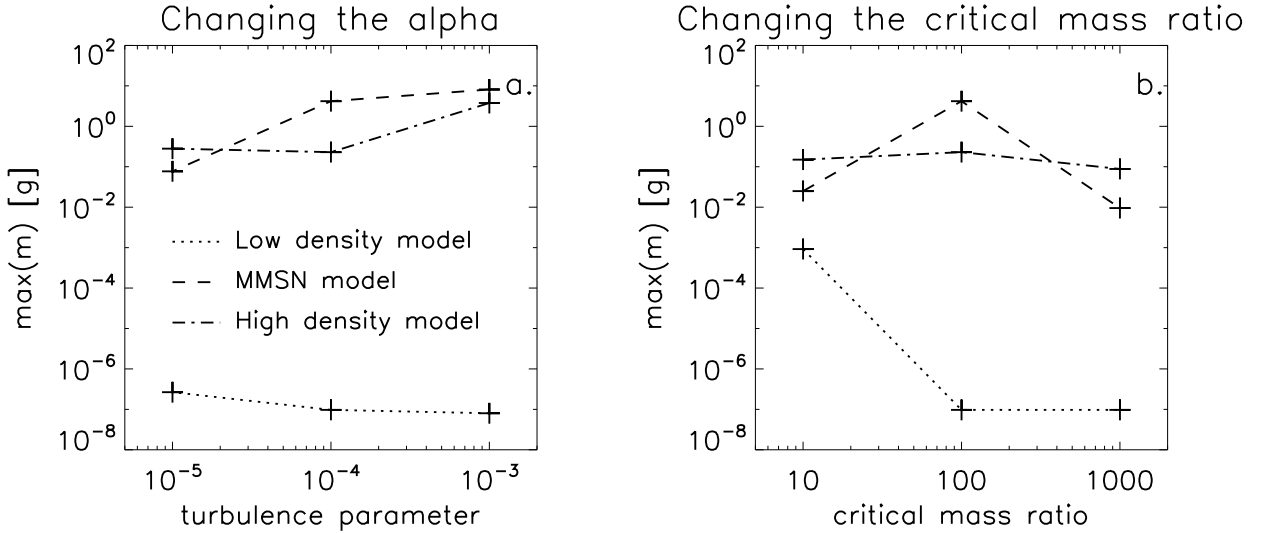}
  \caption{The maximum mean particle mass as a function of the turbulence parameter (a) and critical mass ratio (b).}
  \label{fig:res}
\end{figure*}

To explore the effects of turbulence, we perform two more simulations in each of the disk models. We keep the critical mass ratio fixed (100) and vary only the turbulence parameter ($\alpha$) to have values of $10^{-3}$, $10^{-4}$ and $10^{-5}$. The results are shown in Table \ref{tab:res}, the first nine models and in Fig. \ref{fig:res}a. 

The work of \cite{Brauer2008a} suggests that in situations where fragmentation limits the growth, a lower turbulence strength results in bigger aggregates. This, of course, directly reflects the shift of the fragmentation threshold (1 m/s) to larges sizes when $\alpha$ is lower (Fig. \ref{relcont}). In this study it is fragmentation that balances the growth, which results in a (quasi) steady-state. For the low density models we do see a decrease of the final particle mass, but it is bouncing that balances it. In the low density model, particles grow only in the {\Sa} regimes. When particles leave these regimes, the growth stops due to bouncing. The border of the S1 regime is determined by the collision energy being lower than $5\times E_{\mathrm{roll}}$, where $E_{\mathrm{roll}}$ is the rolling energy of monomers (see Paper I). As the collision energy is $E_{\mathrm{coll}} = 1/2 \mu (\Delta v)^2$, particles in strong turbulence leave the S1 regimes at lower particle masses.

On the other hand, the MMSN and high density models show that the maximum mass of the particles can even increase with $\alpha$. The precise value of the $\max (\bar{m})$ is determined by the intensity of the peak in the mass-density plots (Sect. \ref{subsec:mmsn_early}) and this may vary somewhat between the simulations. In the `Mt1d-4m100' model we have argued that the spike is exceptionally pronounced due to the high probability of {\Sb} collisions at the initial part of the fast growth. However the main point is that in the MMSN/high density simulations the maximum particle masses all end up around 1 g, independent of the turbulent strength.

The reason for this is the nature of the S2-B2 transition, which occurs at projectile masses of $10^{-4}$ g in the {\pP} plot. As explained before, collisions in the {\pP} plot are the only way by which particles can grow after the {\Sa} phase is finished. Thus, we require a broad distribution for a high growth rate. However, B2 collisions works in the opposite way: it transfers mass from the target to the projectile, narrowing the distribution and decreasing the overall probability for the {\pP} process. Thus, once B2 becomes effective, there is a shift from the {\pP} panel to the {\pp} panel. For the MMSN/high density models this behavior is always present and the important quantities involved (i.e., relative probability of B2 over S2) scale with mass and not with velocity. The result is that the maximum masses particles achieve are $\sim$ 1 g and rather insensitive to the strength of the turbulence.

\begin{figure*}
\centering
  \includegraphics[width=0.9\textwidth]{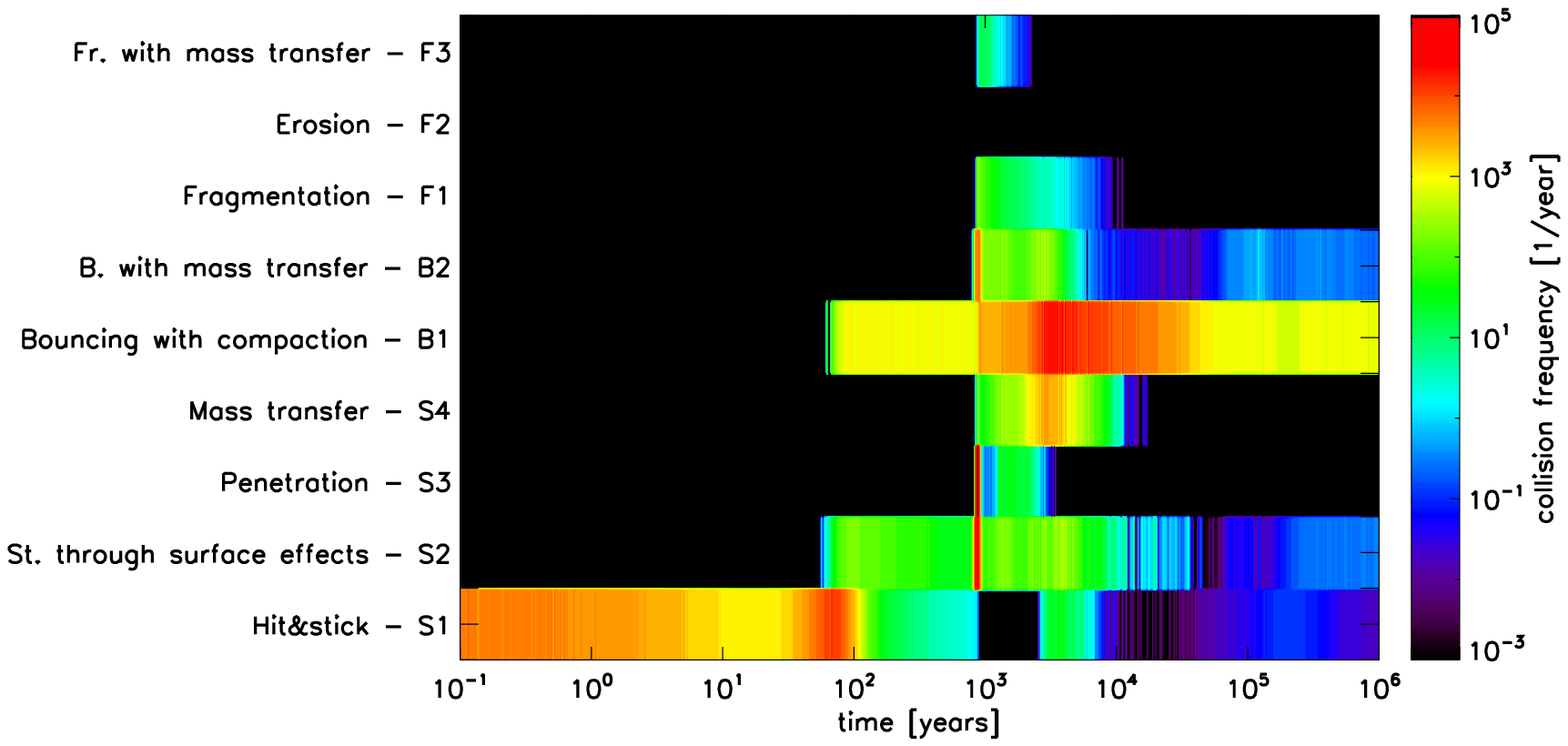}
  \caption{The collision frequencies of the 9 collision types in the MMSN model with $\alpha = 10^{-3}$ and $r_{\mathrm{m}}=100$.}
  \label{fig:mmsn_1d-3}
\end{figure*}

\subsection{Varying the critical mass ratio}
\label{sec:critmass}
We perform simulations in the disk models with $\alpha=10^{-4}$ but with a varying critical mass ratio. We explore how the dust distributions change upon using $r_{\mathrm{m}} = 10$, 100 and 1000. Table \ref{tab:res}, lines 10 to 18, shows the parameters describing the distribution functions, and Fig. \ref{fig:res}b illustrates the maximum particle mass as a function of the critical mass ratio. \\

By examining Table \ref{tab:res} we see that using $r_{\mathrm{m}} = 10$ in the low density model (`Lt1d-4m10') results in heavier and more compact particles. The low critical mass ratio means that the biggest particles in the different sized regimes can sweep up the projectiles and grow to bigger sizes, eventually reaching the fragmentation line, where growth stops. As discussed in Sect. \ref{sec:vrel}, assuming a fragmentation velocity of 1 m s$^{-1}$, the maximum Stokes number of the aggregates is $4.7\times 10^{-3}$. This value is almost reached in this model. 

We find that there is no significant difference between the $r_{\mathrm{m}} = 100$ and 1000 simulations in the low density model. The explanation for this can be found by examining the width of the mass distribution in the {\Sa} phase. This initial phase is happening in the same way independently of the critical mass ratio. If the critical mass ratio $r_m$ is equal to or larger than the width of the distribution function, collisions between different size particles in the {\pP} regime are inhibited. After the S1 phase, the width of the distribution in the low density regime is approximately 100. Therefore, we do not see any difference when the mass threshold is shifted from $r_m$ = 100 to $r_m$ = 1000; in both cases collisions occur between equal-size particles only and these are either S1 or (when this stage is over) B1.

For the high density models (MMSN/Desch) we find that the outcome is again similar: growth halts at $\sim$ 0.1 g (within a factor of 10) and no clear dependence on $r_m$ is seen. For the high mass ratios, growth is always in the similar-size regime. Here, it is the gas density that determines the velocity, i.e., whether we have a sticking (S1) or a bouncing (B1) collision. Therefore, if $r_m=1000$, the high density model produces heavier particles than the MMSN model (see Fig. \ref{fig:res}b). For lower $r_m$ it is again the nature of the S2-B2 transition regime that limits the maximum mass.

Thus, the critical mass ratio is an important parameter since it determines the relative likelihood of collisions occurring in the different-size regime, which are in general more conducive to growth. Conversely, in simulations where B2 collisions are important -- which have the effect to narrow the distribution -- the width of the distribution will correspond to the value of the $r_m$ parameter, although we have also seen that the absolute size/mass is rather insensitive to it. Overall, these arguments indicate that a good knowledge of this parameter is important.

\section{Discussion}
\label{sect:disc}
We performed simulations with varying turbulence parameter and critical mass ratio values in three disk models having low, intermediate and high gas densities. We find that {\Sa} and {\Ba} are the most dominant collision types. All simulations show the presence of long lived, quasi-steady states. Fragmentation is rarely present, but even then, only for a limited time period. The absence of fragmentation is due to the bouncing collisions.

\subsection{The sensitivity of the results}
\label{subsec:sens}
As presented in Sect. \ref{sec:res}, the outcome of our simulations is determined by the collision kernel, the relative velocity field. A significant change in one, or both can alter the evolution of the aggregates. 

Here we present the results of a test simulation, where the {\Sb} -- {\Bb} transition regime in the {\pP} plot is neglected and replaced by S2 collisions. This alternative transition regime provides a good opportunity to further examine the fast growth presented in Sect. \ref{subsec:mmsn_early}, as the kernel is now simplified. The new kernel also gives us the possibility to see how much the outcome of our simulations can be altered by changing critical areas of the parameter space. As the transition regime is only constrained by one experiment in a rather small area (see e.g. Fig. \ref{lowd_cont} or Fig. 11 in Paper I), further experiments may make it necessary to change this part of the parameter space. We use the same initial conditions as in the `Mt1d-4m100' model described in Sect. \ref{sec:mmsn}. 

In this case, the heaviest particle experiences increased relative velocities, as soon as it reaches $m=0.1$ g, and the particle undergoes a fast growth period (as in the original MMSN simulation, Sect. \ref{sec:mmsn}). Figure \ref{fig:acc} illustrates the growth timescale of the heaviest particle (dotted line) and the minimum growth timescale possible (solid line). As there are no B2 collisions to reduce the mass of the heaviest particle, the growth timescale reaches the maximum that is possible. The heaviest particle increases in mass until the rest of the particle population enters the B2 regimes above 0.1 g in the {\pP} plot. In this simulation, the maximum average mass is 27 g, whereas in the original simulation with the transition regime, the value is 4.18 g.

This work, together with Paper I, is the first attempt to calculate dust growth in protoplanetary disks on an empirical, thus more realistic basis. However, a few more cycles of the feedback loop between the laboratory experiments, the models of the kind described by Paper I and the models described in the paper have to be conducted before we can get near a truly reliable model of dust growth in protoplanetary disks. 

\begin{figure}
\centering
  \includegraphics[width=0.5\textwidth]{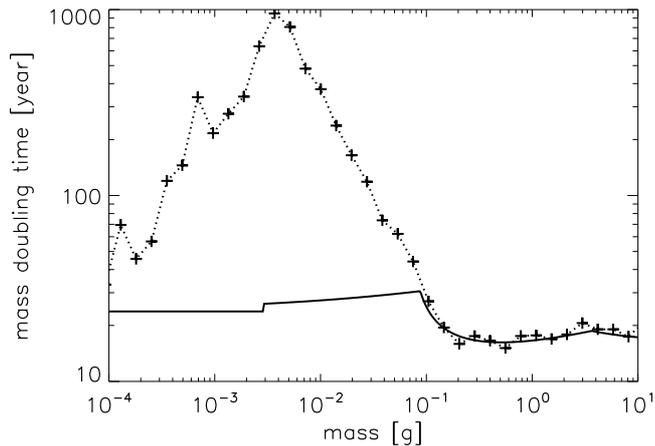}
  \caption{Growth timescale in the test simulation where the S2-B2 transition regime is replaced by S2 collisions only. The dotted line represents the growth timescale of the heaviest particles, the solid line is the minimum growth timescale. In this scenario, the growth timescale reaches the maximum possible value.}
  \label{fig:acc}
\end{figure}

\subsection{Retention of small grains}
\cite{Dullemond:2005p78} showed that without a mechanism that reduces the sticking probability of particles in the upper layers of the disk or without a continuous source of small particles, the observed SEDs of TTauri stars would show very weak infrared excess. The SEDs of TTauri stars have strong IR excess (e.g. \cite{Furlan2005}, \cite{Kessler-Silacci2006}); therefore, some kind of grain-retention mechanism is needed to explain these SEDs. Previous models of grain growth assumed a continuous cycle of growth and fragmentation, which provides the necessary amount of small particles (see e.g. \cite{Brauer2008a}, \cite{Dullemond:2005p78}, Birnstiel et al. (2009)). Our simulations, however, showed that the mass distribution function is narrow. Small, monomer sized particles are not present and fragmentation is ineffective in providing small particles, which could be transported to disk atmospheres. The question naturally arises: how can small grains be produced in our collision model? \\

One possible solution might come from bouncing. \cite{Weidling2009} performed bouncing experiments by putting an aggregate onto an oscillating metal plate and measuring the porosity of particles due to collisions with the plate. They observed that approximately 10\% of the projectile mass eroded during the experiment (see Table 1 of their paper). This mass loss can happen due to the initial collisions; thus the eroded mass sticked to the baseplate. It is also possible that small pieces of fragments grind off when the aggregates bounce, which cannot be observed in the experiment. These ground off particles can then diffuse out of the midplane and provide the necessary amount of small particles to the upper layers of the disk. Future laboratory experiments are needed to quantify the level of ground off particles in bouncing collisions.\\

The second possible explanation is provided by dust growth at the upper layers of the disk. We performed two simulations at four pressure scale-heights in the low density model and in the MMSN model using $\alpha=10^{-4}$. We find that the relative velocity of two monomers in the Brauer model is 2 m s$^{-1}$, thus monomers at these heights do not coagulate, only bounce. The particles in the MMSN model can form aggregates of maximum of 10 $\mu$m in size. Using a higher $\alpha$ (as is mostly assumed in the upper layers of the disk) can completely halt even this limited growth. Therefore, bouncing could be the key ingredient the mechanism that reduces the sticking probability of the particles. However, if substantial vertical turbulent mixing takes place, this may not help, because these monomers would then be ``vacuum cleaned'' away by the bigger particles at the interior of the disk. Further studies of 1D vertical slices of disk models are needed to investigate this scenario. 

\subsection{Implications for planetesimal formation models}
\label{subsec:plf}
One can also see that coagulation only cannot produce planetesimals with the conditions presented in this work. Even if the turbulence parameter is taken to be zero, relative velocity due to radial drift is preventing particles to cross the so called 'meter size barrier'. An ideal environment for particle growth is a pressure bump in the dead zone where both the turbulent and radial relative velocities are reduced. Such an environment is located around the snow line (\cite{Kretke2007}). \cite{Brauer2008b} showed that in these pressure bumps relative velocities stayed below a presumed fragmentation threshold of 10 m s$^{-1}$, presenting a window through which particles can overcome the m-size barrier, although they assumed perfect sticking (no bouncing) below the fragmentation barrier. Future studies have to verify whether planetesimals can be formed with the collision model presented in this study.\\

Another planetesimal forming mechanism is the gravitational collapse of swarms of boulders (\cite{Johansen:2007p65}). This scenario assumes that large amount of the solid material is presented in dm sized boulders ($St \ge 0.1$) at the midplane of the disk. These boulders then concentrate in long-lived high pressure regions in the turbulent gas and these initial over-densities are further amplified by the streaming instability. This mechanism forms 100 km sized objects on a very short timescale (some orbits). However, our simulations produce particles with $St \approx 10^{-4}$ which is due to {\Ba} and the low (1 m s$^{-1}$) fragmentation velocity of silicates. Using a 'stickier' material such as ices or particles with organic mantels may produce bigger particles. Molecular dynamic simulations (e.g. \cite{Dominik:1997p89}, \cite{Wada2007}, \cite{Wada2008}) showed that icy aggregates could have fragmentation velocities of about 10 m s$^{-1}$, although these findings have yet to be confirmed by laboratory experiments. Similarly, it is conceivable that the enhanced sticking capabilities of ices will prevent the bouncing, which is so omnipresent for small particles in our simulations, or shifts it to larger sizes.\\

\cite{Cuzzi2008} outlined an alternative concentration mechanism to obtain gravitationally unstable clumps of particles, which can then undergo sedimentation and form a `sandpile' planetesimal. In this model turbulence causes dense concentrations of aerodynamically size-sorted, chondrule-size particles (\cite{Cuzzi2001})-- more precisely, particles of Stokes numbers $St=Re^{-1/2}\approx 10^{-4}$ in our simulations. Since growth in our models is typically halted at these Stokes numbers, this concentration mechanism is an obvious successor to coagulation -- at least where it concerns the conditions adopted in this paper (1AU, silicates). \\

However, it should be emphasized that the formation of a gravitationally unstable clump does not imply planetesimals will form unimpededly. An important question to address is how collisions will affect the collapse. In the \cite{Cuzzi2008} scenario the collapse occurs on a sedimentation timescale and for these high densities collisions between particles will be frequent. Likewise, in the Johansen scenario -- where the collapse occurs on an orbital timescale and involves $St \sim 0.1$ particles -- collisions can be rather violent. Collisional fragmentation or erosion may change the appearance of the collapse, because the small fragments are carried away by the gas. The role of collisions in these situations is certainly an important question, and our new collision model provides a tool to quantitatively address this issue in future studies.

\subsection{Consequences for laboratory experiments} 
\label{subsec:lab}
One can see from Fig. 11, in Paper I, that only a small part of the parameter space is covered by experiments. Although laboratory experiments cannot be made at every point of the parameter space, we suggest future ones based on Figs. \ref{lowd_cont}, \ref{mmsn_cont} and \ref{desch_cont} in order to better understand dust growth in the early stages of planet formation.
\begin{itemize}
\item More experiments in the {\cc} and {\cC} regimes are needed as particles get compactified by the end of their evolution. Thus, most of the collisions happen in this regime, at velocities between 0.1 and 100 cm s$^{-1}$, at masses between $10^{-7}$ and 10 g.
\item As seen in Figs. \ref{lowd_cont}, \ref{mmsn_cont} and \ref{desch_cont}, the `hot spots', where most of the collisions are happening, are located in the equal sized regimes, at the left side of the fragmentation line. Therefore, it is important to map these areas of the parameter space in detail.
\item We define a sharp border line between the {\Sa} and {\Ba} collisions. If there is a continuous transition between S1 and B1, the growth of particles would not be halted by bouncing at such low particle sizes. As many collisions are happening in the {\pp} and {\cc} regimes, even a small probability of growth could increase the particle sizes.
\item As seen in Fig. \ref{mmsn_pics}b, particles in high gas density environments can have enlargement parameters much higher than 6.6 ($\phi = 0.15$). An interesting question is whether the collision types and regimes are also valid for particles with such a low volume filling factors, or whether these particles have a different collision behavior?
\item The {\Sb} -- {\Bb} transition regime greatly affects the outcome of the simulations (see Sect. \ref{subsec:sens}). However, the transition regime is only mapped at the high velocity and low mass regions. Therefore, it is essential to better constrain this part of the parameter space.
\item The critical mass ratio affects the particle masses and porosities. Experiments are needed to constrain its value.
\item The bouncing model, described in Paper I, has important implications for the evolution of dust aggregates in protoplanetary disks but it is unfortunately still based on too few experiments. Further experiments are needed to refine the model, as {\Ba} is the most frequent collision type in all of the simulations. 
\end{itemize} 

\section{Summary}
\label{sec:sum}
We performed simulations of dust growth using the Monte Carlo code of ZsD08 and a dust collision model based on laboratory experiments (Paper I). We performed simulations at the midplane of three disk models having low ($2.4 \times 10^{-11}$ g cm$^{-3}$), intermediate ($1.4 \times 10^{-9}$ g cm$^{-3}$) and high ($2.7\times 10^{-8}$ g cm$^{-3}$) gas densities at 1 AU distance from the central star. We vary the turbulence parameter ($\alpha$) and the critical mass ratio ($r_{\mathrm{m}}$) to explore their effects on the mass and porosity distribution functions. Our main results are:
\begin{itemize}
\item Upon using $\alpha = 10^{-4}$, the low density / MMSN / high density model produces particles with maximum mean mass of $9.7 \times 10^{-8}$ g / 4.18 g / 0.23 g, the maximum average enlargement parameter of these particles are 7.12 / 21.9 / 38.0. The maximum average Stokes numbers are $2.2\times 10^{-4}$ / $2.8\times 10^{-4}$ / $3.6 \times 10^{-5}$. 
\item We find that particle evolution does not follow the previously assumed growth-fragmentation cycles. Although catastrophic fragmentation is present for a short period of time in some of the models (typically when $\alpha = 10^{-3}$), it has a fringe effect. Particles in most of the simulations do not reach the fragmentation barrier because their growth is halted by bouncing.
\item We see long lived, quasi-steady states in the distribution function of the aggregates due to bouncing. The final equilibrium state is not reached within the drift or the viscous timescales.
\item We performed simulations with varying turbulence strength. We find that the system is `non-linear': The maximum mass of particles is not a decreasing function of the turbulence parameter and is not an increasing function of the gas density.
\item We explored the effects of the critical mass ratio. We find that different critical mass ratios can affect the particle evolution. Small critical mass ratios can produce heavier particles, while big values of $r_m$ can halt the growth earlier.
\item The maximum Stokes number is rather independent of the gas density and the strength of the turbulence.
\item The maximum mass of the aggregates is limited to $\approx$ 1 g due to the S2-B2 transition regime.
\item The Stokes number $10^{-4}$ particles can be concentrated in turbulence by aerodynamical size-sorting, thus planetesimals can form from these particles.
\end{itemize}
 
\begin{acknowledgement}
A. Zs. thanks to Frithjof Brauer, Tilman Birnstiel and Felipe Gerhard for useful discussions. C.G. was funded by the Deutsche Forschungsgemeinschaft within the Forschergruppe 759 ``The Formation of Planets: The Critical First Growth Phase'' under grant Bl 298/7-1. A. Zs. acknowledges the support of the IMPRS for Astronomy \& Cosmic Physics at the University of Heidelberg and C.W.O. acknowledges the financial support from the Alexander von Humboldt Foundation. We also thank our referee, Sascha Kempf for providing useful comments which helped to improve the paper. 
\end{acknowledgement}

\begin{appendix}
\section{Time evolution and animations}
We offer six animations to illustrate the time evolution of the aggregates. These animations can be viewed at \texttt{www.mpia.de/homes/zsom/Site/animations2.html}. The \texttt{mass\_vs\_psi\_lowd.mpg} file shows the time evolution of the masses and enlargement parameters of our representative particles. The x-axis is the particle mass in gram units, the y-axis is the enlargement parameter. The \texttt{regimes\_lowd.mpg} file illustrates the time evolution of Fig. \ref{lowd_cont}. The contour levels are collision frequencies / grid cell. The same movies are provided for the MMSN model, these are \texttt{mass\_vs\_psi\_mmsn.mpg} and \texttt{regimes\_mmsn.mpg} respectively. And also for the high density model: \texttt{mass\_vs\_psi\_highd.mpg} and \texttt{regimes\_highd.mpg}. 
\end{appendix}

\bibliographystyle{bibtex/aa}

\begin{thebibliography}{60}
\expandafter\ifx\csname natexlab\endcsname\relax\def\natexlab#1{#1}\fi

\bibitem[{{Andrews} \& {Williams}(2007)}]{Andrews2007}
{Andrews}, S.~M. \& {Williams}, J.~P. 2007, \apj, 659, 705

\bibitem[{{Barge} \& {Sommeria}(1995)}]{Barge1995}
{Barge}, P. \& {Sommeria}, J. 1995, \aap, 295, L1

\bibitem[{{Blum} \& {M\"unch}(1993)}]{Blum1993}
{Blum}, J. \& {M\"unch}, M. 1993, Icarus, 106, 151

\bibitem[{{Blum} \& {Schr{\"a}pler}(2004)}]{Blum2004}
{Blum}, J. \& {Schr{\"a}pler}, R. 2004, Physical Review Letters, 93, 115503

\bibitem[{{Blum} \& {Wurm}(2000)}]{Blum2000}
{Blum}, J. \& {Wurm}, G. 2000, Icarus, 143, 138

\bibitem[{{Blum} \& {Wurm}(2008)}]{Blum2008}
{Blum}, J. \& {Wurm}, G. 2008, \araa, 46, 21

\bibitem[{{Blum} {et~al.}(1996){Blum}, {Wurm}, {Kempf}, \&
  {Henning}}]{Blum1996}
{Blum}, J., {Wurm}, G., {Kempf}, S., \& {Henning}, T. 1996, Icarus, 124, 441

\bibitem[{{Blum} {et~al.}(2000){Blum}, {Wurm}, {Kempf}, {Poppe}, {Klahr},
  {Kozasa}, {Rott}, {Henning}, {Dorschner}, {Schr{\"a}pler}, {Keller},
  {Markiewicz}, {Mann}, {Gustafson}, {Giovane}, {Neuhaus}, {Fechtig},
  {Gr{\"u}n}, {Feuerbacher}, {Kochan}, {Ratke}, {El Goresy}, {Morfill},
  {Weidenschilling}, {Schwehm}, {Metzler}, \& {Ip}}]{Blum2000a}
{Blum}, J., {Wurm}, G., {Kempf}, S., {et~al.} 2000, Physical Review Letters,
  85, 2426

\bibitem[{Brauer {et~al.}(2008a)Brauer, Dullemond, \& Henning}]{Brauer2008a}
Brauer, F., Dullemond, C.~P., \& Henning, T. 2008a, A{\&}A, 480, 859

\bibitem[{{Brauer} {et~al.}(2008b){Brauer}, {Henning}, \&
  {Dullemond}}]{Brauer2008b}
{Brauer}, F., {Henning}, T., \& {Dullemond}, C.~P. 2008b, \aap, 487, L1

\bibitem[{{Chambers}(2001)}]{Chambers2001}
{Chambers}, J.~E. 2001, Icarus, 152, 205

\bibitem[{{Cuzzi} {et~al.}(2001){Cuzzi}, {Hogan}, {Paque}, \&
  {Dobrovolskis}}]{Cuzzi2001}
{Cuzzi}, J.~N., {Hogan}, R.~C., {Paque}, J.~M., \& {Dobrovolskis}, A.~R. 2001,
  \apj, 546, 496

\bibitem[{{Cuzzi} {et~al.}(2008){Cuzzi}, {Hogan}, \& {Shariff}}]{Cuzzi2008}
{Cuzzi}, J.~N., {Hogan}, R.~C., \& {Shariff}, K. 2008, \apj, 687, 1432

\bibitem[{{Desch}(2007)}]{Desch2007}
{Desch}, S.~J. 2007, \apj, 671, 878

\bibitem[{Dominik \& Tielens(1997)}]{Dominik:1997p89}
Dominik, C. \& Tielens, A. G. G.~M. 1997, Astrophysical Journal v.480, 480, 647

\bibitem[{Dullemond \& Dominik(2004)}]{Dullemond:2004p325}
Dullemond, C.~P. \& Dominik, C. 2004, A{\&}A, 421, 1075

\bibitem[{Dullemond \& Dominik(2005)}]{Dullemond:2005p78}
Dullemond, C.~P. \& Dominik, C. 2005, A{\&}A, 434, 971

\bibitem[{{Epstein}(1924)}]{Epstein1924}
{Epstein}, P.~S. 1924, Physical Review, 23, 710

\bibitem[{{Furlan} {et~al.}(2005){Furlan}, {Calvet}, {D'Alessio}, {Hartmann},
  {Forrest}, {Watson}, {Uchida}, {Sargent}, {Green}, \& {Herter}}]{Furlan2005}
{Furlan}, E., {Calvet}, N., {D'Alessio}, P., {et~al.} 2005, \apjl, 628, L65

\bibitem[{{Garaud} \& {Lin}(2004)}]{Garaud2004}
{Garaud}, P. \& {Lin}, D.~N.~C. 2004, \apj, 608, 1050

\bibitem[{{Hayashi} {et~al.}(1985){Hayashi}, {Nakazawa}, \&
  {Nakagawa}}]{Hayashi1985}
{Hayashi}, C., {Nakazawa}, K., \& {Nakagawa}, Y. 1985, in Protostars and
  Planets II, ed. D.~C. {Black} \& M.~S. {Matthews}, 1100--1153

\bibitem[{Johansen {et~al.}(2007)Johansen, Oishi, Low, Klahr, Henning, \&
  Youdin}]{Johansen:2007p65}
Johansen, A., Oishi, J.~S., Low, M.-M.~M., {et~al.} 2007, Nature, 448, 1022

\bibitem[{{Kempf} {et~al.}(1999){Kempf}, {Pfalzner}, \& {Henning}}]{Kempf1999}
{Kempf}, S., {Pfalzner}, S., \& {Henning}, T.~K. 1999, Icarus, 141, 388

\bibitem[{{Kessler-Silacci} {et~al.}(2006){Kessler-Silacci}, {Augereau},
  {Dullemond}, {Geers}, {Lahuis}, {Evans}, {van Dishoeck}, {Blake}, {Boogert},
  {Brown}, {J{\o}rgensen}, {Knez}, \& {Pontoppidan}}]{Kessler-Silacci2006}
{Kessler-Silacci}, J., {Augereau}, J.-C., {Dullemond}, C.~P., {et~al.} 2006,
  \apj, 639, 275

\bibitem[{{Klahr} \& {Henning}(1997)}]{Klahr1997}
{Klahr}, H.~H. \& {Henning}, T. 1997, Icarus, 128, 213

\bibitem[{{Kokubo} {et~al.}(2006){Kokubo}, {Kominami}, \& {Ida}}]{Kokubo2006}
{Kokubo}, E., {Kominami}, J., \& {Ida}, S. 2006, \apj, 642, 1131

\bibitem[{{Kornet} {et~al.}(2001){Kornet}, {Stepinski}, \&
  {R{\'o}{\.z}yczka}}]{Kornet2001}
{Kornet}, K., {Stepinski}, T.~F., \& {R{\'o}{\.z}yczka}, M. 2001, \aap, 378,
  180

\bibitem[{{Krause} \& {Blum}(2004)}]{Krause2004}
{Krause}, M. \& {Blum}, J. 2004, Physical Review Letters, 93, 021103

\bibitem[{Kretke \& Lin(2007)}]{Kretke2007}
Kretke, K.~A. \& Lin, D. N.~C. 2007, The Astrophysical Journal, 664, L55

\bibitem[{{Lyra} {et~al.}(2009){Lyra}, {Johansen}, {Zsom}, {Klahr}, \&
  {Piskunov}}]{Lyra2009a}
{Lyra}, W., {Johansen}, A., {Zsom}, A., {Klahr}, H., \& {Piskunov}, N. 2009,
  \aap, 497, 869

\bibitem[{{Markiewicz} {et~al.}(1991){Markiewicz}, {Mizuno}, \&
  {Voelk}}]{Markiewicz1991}
{Markiewicz}, W.~J., {Mizuno}, H., \& {Voelk}, H.~J. 1991, \aap, 242, 286

\bibitem[{{Mizuno}(1980)}]{Mizuno1980}
{Mizuno}, H. 1980, Progress of Theoretical Physics, 64, 544

\bibitem[{{Mizuno} {et~al.}(1988){Mizuno}, {Markiewicz}, \&
  {Voelk}}]{Mizuno1988}
{Mizuno}, H., {Markiewicz}, W.~J., \& {Voelk}, H.~J. 1988, \aap, 195, 183

\bibitem[{{Nakagawa} {et~al.}(1983){Nakagawa}, {Hayashi}, \&
  {Nakazawa}}]{Nakagawa1983}
{Nakagawa}, Y., {Hayashi}, C., \& {Nakazawa}, K. 1983, Icarus, 54, 361

\bibitem[{{Nomura} \& {Nakagawa}(2006)}]{Nomura2006}
{Nomura}, H. \& {Nakagawa}, Y. 2006, \apj, 640, 1099

\bibitem[{{Okuzumi}(2009)}]{Okuzumi2009}
{Okuzumi}, S. 2009, \apj, 698, 1122

\bibitem[{Ormel \& Cuzzi(2007)}]{Ormel:2007p92}
Ormel, C.~W. \& Cuzzi, J.~N. 2007, A{\&}A, 466, 413

\bibitem[{Ormel \& Spaans(2008)}]{Ormel:2008p95}
Ormel, C.~W. \& Spaans, M. 2008, The Astrophysical Journal, 684, 1291

\bibitem[{Ormel {et~al.}(2007)Ormel, Spaans, \& Tielens}]{Ormel:2007p93}
Ormel, C.~W., Spaans, M., \& Tielens, A. G. G.~M. 2007, A{\&}A, 461, 215

\bibitem[{Ossenkopf(1993)}]{Ossenkopf:1993p80}
Ossenkopf, V. 1993, Astronomy and Astrophysics (ISSN 0004-6361), 280, 617

\bibitem[{{Pollack} {et~al.}(1996){Pollack}, {Hubickyj}, {Bodenheimer},
  {Lissauer}, {Podolak}, \& {Greenzweig}}]{Pollack1996}
{Pollack}, J.~B., {Hubickyj}, O., {Bodenheimer}, P., {et~al.} 1996, Icarus,
  124, 62

\bibitem[{{Safronov}(1969)}]{Safronov1969}
{Safronov}, V.~S. 1969, {Evolution of the Protoplanetary Cloud and Formation of
  the Earth and Planets} (Moscow: Nauka; English transl. NASA TTF-677 [1972])

\bibitem[{{Schmitt} {et~al.}(1997){Schmitt}, {Henning}, \&
  {Mucha}}]{Schmitt1997}
{Schmitt}, W., {Henning}, T., \& {Mucha}, R. 1997, \aap, 325, 569

\bibitem[{{Shakura} \& {Sunyaev}(1973)}]{Shakura1973}
{Shakura}, N.~I. \& {Sunyaev}, R.~A. 1973, \aap, 24, 337

\bibitem[{{Suyama} {et~al.}(2008){Suyama}, {Wada}, \& {Tanaka}}]{Suyama2008}
{Suyama}, T., {Wada}, K., \& {Tanaka}, H. 2008, \apj, 684, 1310

\bibitem[{{Takeuchi} \& {Lin}(2002)}]{Takeuchi2002}
{Takeuchi}, T. \& {Lin}, D.~N.~C. 2002, \apj, 581, 1344

\bibitem[{{Tanaka} {et~al.}(2005){Tanaka}, {Himeno}, \& {Ida}}]{Tanaka2005}
{Tanaka}, H., {Himeno}, Y., \& {Ida}, S. 2005, \apj, 625, 414

\bibitem[{{Thommes} {et~al.}(2008){Thommes}, {Matsumura}, \&
  {Rasio}}]{Thommes2008}
{Thommes}, E.~W., {Matsumura}, S., \& {Rasio}, F.~A. 2008, Science, 321, 814

\bibitem[{{Tsiganis} {et~al.}(2005){Tsiganis}, {Gomes}, {Morbidelli}, \&
  {Levison}}]{Tsiganis2005}
{Tsiganis}, K., {Gomes}, R., {Morbidelli}, A., \& {Levison}, H.~F. 2005, \nat,
  435, 459

\bibitem[{{V\"{o}lk} {et~al.}(1980){V\"{o}lk}, {Jones}, {Morfill}, \&
  {Roeser}}]{Voelk1980}
{V\"{o}lk}, H.~J., {Jones}, F.~C., {Morfill}, G.~E., \& {Roeser}, S. 1980,
  \aap, 85, 316

\bibitem[{{Wada} {et~al.}(2007){Wada}, {Tanaka}, {Suyama}, {Kimura}, \&
  {Yamamoto}}]{Wada2007}
{Wada}, K., {Tanaka}, H., {Suyama}, T., {Kimura}, H., \& {Yamamoto}, T. 2007,
  \apj, 661, 320

\bibitem[{{Wada} {et~al.}(2008){Wada}, {Tanaka}, {Suyama}, {Kimura}, \&
  {Yamamoto}}]{Wada2008}
{Wada}, K., {Tanaka}, H., {Suyama}, T., {Kimura}, H., \& {Yamamoto}, T. 2008,
  \apj, 677, 1296

\bibitem[{{Weidenschilling}(1977{\natexlab{a}})}]{Weidenschilling1977}
{Weidenschilling}, S.~J. 1977{\natexlab{a}}, \mnras, 180, 57

\bibitem[{{Weidenschilling}(1977{\natexlab{b}})}]{Weidenschilling1977a}
{Weidenschilling}, S.~J. 1977{\natexlab{b}}, \apss, 51, 153

\bibitem[{{Weidenschilling}(1980)}]{Weidenschilling1980}
{Weidenschilling}, S.~J. 1980, Icarus, 44, 172

\bibitem[{{Weidenschilling} \& {Cuzzi}(1993)}]{Weidenschilling1993}
{Weidenschilling}, S.~J. \& {Cuzzi}, J.~N. 1993, in Protostars and Planets III,
  ed. {E.~H.~Levy \& J.~I.~Lunine}, 1031--1060

\bibitem[{{Weidling} {et~al.}(2009){Weidling}, {G{\"u}ttler}, {Blum}, \&
  {Brauer}}]{Weidling2009}
{Weidling}, R., {G{\"u}ttler}, C., {Blum}, J., \& {Brauer}, F. 2009, \apj, 696,
  2036

\bibitem[{Wetherill(1990)}]{Wetherill:1990p85}
Wetherill, G.~W. 1990, Icarus (ISSN 0019-1035), 88, 336

\bibitem[{{Whipple}(1972)}]{Whipple1972}
{Whipple}, F.~L. 1972, in From Plasma to Planet, ed. A.~{Elvius}, 211

\bibitem[{{Zsom} \& {Dullemond}(2008)}]{Zsom2008}
{Zsom}, A. \& {Dullemond}, C.~P. 2008, \aap, 489, 931

\end{thebibliography}

\end{document}